\documentclass[a4paper,12pt]{article}
\pdfoutput=1
\usepackage{graphicx}
\usepackage{epsf, amssymb,amsmath,dsfont}
\usepackage{epsfig}
\usepackage{hyperref}
\usepackage{color}
\usepackage{mathtools,float}
\usepackage{array} 

\numberwithin{equation}{section}

\definecolor{dark-blue}{rgb}{0.15,0.15,0.4}
\hypersetup{
    colorlinks
    ,linkcolor={black},
    citecolor={blue}
}

\makeatletter
\renewcommand\section{\@startsection {section}{1}{\z@}%
                                   {-3.5ex \@plus -1ex \@minus -.2ex}
                                   {2.3ex \@plus.2ex}%
                                   {\normalfont\large\bfseries}}

\renewcommand\subsection{\@startsection{subsection}{2}{\z@}%
                                     {-3.25ex\@plus -1ex \@minus -.2ex}%
                                     {1.5ex \@plus .2ex}%
                                     {\normalfont\bfseries}}

\usepackage[normalem]{ulem}
\newcommand\out{\bgroup\markoverwith
{\textcolor{red}{\rule[.5ex]{2pt}{1.5pt}}}\ULon}
\usepackage{xcolor}

\makeatother

\newcommand{\bea}{\begin{eqnarray}}
\newcommand{\eea}{\end{eqnarray}}
\newcommand{\be}{\begin{equation}}
\newcommand{\ee}{\end{equation}}
\newcommand{\bem}{\begin{pmatrix}}
\newcommand{\eem}{\end{pmatrix}}
\newcommand{\bl}{\begin{align}}
\newcommand{\el}{\end{align}}

\def \textAdS {{AdS}}

\def \textdS {{dS}}

\def \textMink {{Mink}}

\def \textBTZ {{BTZ}}

\def \textCFT {{CFT}}

\begin{document}

\begin{center} 
 
 ${}$ 
 
\thispagestyle{empty}




\vskip  .1cm {\LARGE {\bf Towards non-AdS Holography  \\[4mm] 
via  the Long String Phenomenon   }}

\vskip  1.2cm {   {Sam van Leuven\footnote{ s.p.g.vanleuven@uva.nl}, \,    Erik Verlinde\footnote{
e.p.verlinde@uva.nl} \, \text{and} \,   Manus Visser\footnote{ m.r.visser@uva.nl} }}\\
{\vskip 1cm   { Institute of Physics and \\ Delta Institute for Theoretical Physics \\ 
University of Amsterdam\\ 
Science Park 904, 
1090 GL Amsterdam\\ The Netherlands} }

\vspace{1.2cm}

\date{\today}

\begin{abstract}
\baselineskip=16pt

The microscopic description of AdS space obeys the holographic principle in the sense that the number of microscopic degrees of freedom is given by the area of the holographic boundary. We assume the same applies to the  microscopic  holographic theories for  non-AdS spacetimes, specifically for Minkowski, de Sitter, and AdS below its curvature radius. By taking general lessons from   AdS/CFT   we derive the cut-off energy of the holographic theories for these non-AdS geometries.    Contrary to AdS/CFT, the excitation energy decreases towards the IR in the bulk, which is related to the negative specific heat of black holes. We construct a conformal mapping between the non-AdS geometries and $AdS_3\!\times\! S^{q}$ spacetimes, and relate the microscopic properties of the holographic theories for non-AdS spaces to those of     symmetric product CFTs.  We find that the mechanism responsible for the inversion of the energy-distance relation corresponds to the long string phenomenon. This same mechanism naturally explains the negative specific heat for non-AdS black holes and the value of the  vacuum energy in (A)dS spacetimes.  
${}$
\end{abstract}

\end{center}
\newpage

\tableofcontents

\newpage

\section{Introduction}

In the last 20 years considerable progress has been made on the holographic description~of anti-de Sitter space \cite{Maldacena:1997re,Witten:1998qj,Aharony:1999ti}. An important open question concerns the reconstruction of the~bulk spacetime from  the   boundary theory. In addition, little is known about the microscopic theories for more~general~spacetimes, such as Minkowski or de Sitter space. 

One of the general lessons from the AdS/CFT correspondence is the concept of holographic renormalization \cite{Henningson:1998gx, Akhmedov:1998vf,Alvarez:1998wr, Balasubramanian:1999jd, Skenderis:1999mm,deBoer:1999tgo}.  If one moves  the holographic boundary into the bulk one introduces a UV cut-off in the conformal field theory. Going further into the bulk increases  the cut-off length scale and reduces the number of microscopic degrees of freedom of the holographic theory.  This  holographic RG description of AdS spacetimes works well near the boundary at scales large compared to the curvature radius. The microscopic theory is, however, not so well understood if the boundary reaches the AdS scale and breaks down at sub-AdS scales.  Here the curvature radius of AdS is irrelevant, and the geometry becomes approximately that of flat space.  

A crucial hint about  the microscopic theory for non-AdS geometries comes from the holographic principle, motivated by  the Bekenstein-Hawking entropy formula \cite{Bekenstein:1973ur,Hawking:1974sw}:
\\[-2mm]
\begin{equation}  
S= {A\over 4G}   \label{BHformula} \, .
\end{equation}
The holographic principle states that the maximal number of microscopic degrees of freedom associated to a spacelike region is proportional to the area $A$ of its boundary in Planckian units  \cite{tHooft:1993dmi, Susskind:1994vu,Bousso:1999xy,Bousso:2002ju}. However,
an important indication that the microscopic  
holographic descriptions for super-AdS scales and those for sub-AdS scale and dS and 
Minkowski spacetimes are qualitatively different is given by the properties of black holes in these geometries:
large AdS black holes are known to have a positive specific heat, whereas the specific heat of black holes in sub-AdS, flat or dS is negative.  A complete understanding of this feature from a microscopic holographic perspective is still lacking, which is illustrative of our poor understanding of holography for non-AdS spacetimes. 

In this paper we will   make   modest steps towards answering some of these questions. Our strategy is to follow the same line of reasoning as in the original paper \cite{Susskind:1998dq}, which clarified the holographic nature of the AdS/CFT correspondence by showing that it obeys all the general properties which are expected to hold in a microscopic theory that satisfies the holographic principle.  In their work Susskind and Witten established the UV-IR correspondence that is underlying AdS/CFT by relating the UV cut-off of the microscopic theory to the IR cut-off in the bulk. They furthermore showed that the number of degrees of freedom of the cut-off CFT is proportional to the area of the holographic surface  in Planckian units. They also pointed out that if the temperature of the CFT approaches the cut-off scale all the microscopic degrees of freedom become excited  and produce a state whose entropy is given by the Bekenstein-Hawking entropy for a black hole horizon  which coincides with the holographic boundary.

Following this same logic we focus on general features of the holographic theory for sub-AdS geometries and Minkowski and de Sitter space,  such as the number of microscopic degrees of freedom and the typical energy that is required to excite these degrees of freedom.  We will assume that   these   non-AdS geometries are also described by an underlying microscopic quantum theory, that obeys the general principles of holography.  
 A logical assumption is that the number of microscopic degrees of freedom of the cut-off boundary theory for general spacetimes is also determined by the area of the holographic boundary in Planckian units.  Here, by a `cut-off boundary' we mean a holographic screen located inside the spacetime at a finite distance from its `center'.  In this paper we will for definiteness and simplicity only consider  spherically symmetric spacetimes, so that we can choose the center at the origin.  Our main cases of interest are empty (A)dS and Minkowksi space, but we will also study  Schwarzschild geometries. 

  In addition to the number of degrees of freedom we are interested in the excitation energy per degree of freedom. In sub-AdS, flat and de Sitter space we find that this excitation energy decreases with the distance from the center. One of our main conclusions is that the UV-IR correspondence, familiar from AdS/CFT,  is inverted in these spacetimes: long distances (=IR) in the bulk   correspond  to   low energies  (=IR)   in the microscopic theory. And contrary to AdS/CFT the number of degrees of freedom increases towards the IR of the microscopic theory. Hence, we are dealing with a  holographic quantum theory whose typical excitation energy decreases if the number of degrees of freedom increases.  This fact is directly related to the negative specific heat of black~holes. 

   Our aim is to find an explanation of this counter-intuitive feature of the microscopic theory. 
  For this purpose we employ a conformal map between  three non-AdS geometries and spacetimes of the form $AdS_3\!\times\! S^{q}$.  This conformal map relates general features of the microscopic theories on holographic screens in both spacetimes, and allows us to identify the mechanism responsible for the inversion of the energy-distance relation compared to AdS/CFT.  We find that it is a familiar mechanism, often invoked in the microscopic description of black holes \cite{Strominger:1996sh, Maldacena:1996ds, Dijkgraaf:1996xw}, known as the `long string phenomenon'. This mechanism operates on large symmetric product CFTs, and identifies a twisted sector consisting of `long strings' whose typical excitation energy is considerably smaller than that of the untwisted sector. Our conclusion is that this same long string mechanism reduces the excitation energy at large distances in the bulk and towards the IR of the microscopic theory, and therefore explains the negative specific heat of non-AdS black holes.  Furthermore, it clarifies the   value of the vacuum energy of (A)dS, which, contrary to most expectations, differs from its natural value set by the Planck scale. 

The outline of the  paper is as follows.
 In section \ref{sec:lessons} we use lessons from AdS/CFT to give a geometric definition of the number of holographic degrees of freedom and their excitation energy. In section \ref{sec:towardsholographynonads} we present a conjecture relating the microscopic theories for two Weyl equivalent spacetimes and describe the conformal map from sub-AdS, Minkowski and de Sitter space to $AdS_3\!\times\! S^{q}$ type geometries. Section \ref{sec:longstring}  describes the long string mechanism and its relevance for non-AdS holography. Finally, in section \ref{sec:physicalimplications} we discuss the negative specific heat of black holes and   vacuum energy of  (A)dS spacetimes.

\section{Lessons from the AdS/CFT correspondence}
\label{sec:lessons}

 In this paper we are interested in obtaining a better understanding of the microscopic description of  spacetime. For definiteness we consider $d$-dimensional static, spherically symmetric spacetimes with a metric of the form
\begin{equation} \label{metric}
ds^2 = -  f(R) dt^2 + \frac{d R^2}{f(R)} + R^2 d \Omega_{d-2}^2 \, ,
\end{equation}
where $d\Omega_{d-2}$ is the line element on a $(d-2)$-dimensional unit sphere.
This class of metrics allows us to study (anti-)de Sitter space, flat space and (A)dS-Schwarzschild solutions.  In these geometries we consider a $(d-2)$-surface $\mathcal S$  located at a finite radius $R$, corresponding to the boundary of a spacelike ball-shaped region $\mathcal B$ centered around the origin.  We will call $\mathcal S$ the `holographic screen' or `holographic surface'.   We will study     general features of the microscopic description of these spacetimes, where we imagine that   the quantum theory lives on the holographic screen $\mathcal S$.   We will start with discussing the familiar case of anti-de Sitter space, where we have a good qualitative understanding of the holographic theory, on boundaries at a finite radius $R$.

\subsection{General features of the microscopic holographic theory}
\label{sec:generalfeatures}

Our first goal is to present a number of general features of the holographic description of asymptotically AdS spacetimes in a way that is generalizable to other  spacetimes. 
 Motivated by \cite{Susskind:1998dq}, we focus on the following aspects of the microscopic holographic ~description:\footnote{ The quantities $\mathcal C$ and $1/\epsilon$ correspond to $N_{dof}$ and $\delta$ in \cite{Susskind:1998dq}.   
 } 
\begin{eqnarray}
  \mathcal C &=&  \text{number of  UV  degrees of freedom of the holographic theory}  \nonumber \\
   \epsilon &=& \text{excitation energy per UV degree of freedom}  \nonumber \\
  \mathcal N &=& \text{total energy measured in terms of the cut-off energy $\epsilon$}. \nonumber 
\end{eqnarray}  
 More precisely, by $\epsilon$ we mean the total energy of the maximally excited state divided by the number of UV degrees of freedom. Below we will give a definition of each of these quantities purely in terms of the geometry in the neighbourhood of the holographic surface.  We will motivate and verify these definitions for the case of AdS/CFT, but afterwards we will apply those same definitions to other geometric situations.

The number of degrees of freedom $\mathcal C$ is in the case of AdS/CFT directly related to the central charge of the CFT.  According to the holographic dictionary the central charge $c$ of a CFT dual to Einstein gravity is given by \cite{Myers:2010tj}
 \begin{equation} \label{centralcharge}
\frac{c}{12}=\frac{A(L)}{16\pi G_d}  \qquad \text{with} \qquad A(L) = \Omega_{d-2} L^{d-2} \, ,
\end{equation}
where $L$ is the AdS radius and $G_d$ is Newton's constant in $d$ dimensions. The central charge $c$ is defined in terms of the normalization of the 2-point function of the stress tensor \cite{Osborn:1993cr}.  It measures the number of field theoretic degrees of freedom of the CFT. The central charge $c$ is normalized so that it coincides with the standard  central charge in  2d CFT. In three dimensions it reduces  to the Brown-Henneaux formula \cite{Brown:1986nw}. 

The number of degrees of freedom of the microscopic   theory that lives on a holographic screen $\mathcal S$ at radius $R$ is   given by
\begin{equation} \label{nodof}
\mathcal{C}=\frac{c}{12} \left ( \frac{R}{L}  \right)^{\! \!  d-2} \!\!  =\frac{A(R)}{16\pi G_d} \qquad \text{with} \qquad A(R) = \Omega_{d-2} R^{d-2} \, .
\end{equation}
This result can be interpreted as follows. We imagine that the  CFT lives on   $\mathbb R \times S^{d-2}$, where the radius of the sphere is given by $L$ and $\mathbb R$ corresponds to the time $t$. The sphere is now partitioned in cells of size $\delta$, where the lattice cut-off is  related to the radius $R$ 
through the UV-IR correspondence via $\delta = L^2/R$. Hence the number of cells on the sphere is given by: $ \left ( L/\delta \right)^{d-2} = \left ( R/L \right)^{d-2}$.  Further, the factor $c/12$  counts  the number of quantum mechanical degrees of freedom contained in each cell. This simple argument  due to \cite{Susskind:1998dq} holds in any number of dimensions for   generic holographic~CFTs.

The second quantity $\epsilon$  determines the energy that is required to excite one UV degree of freedom and is inversely related to the UV regulator $\delta$ in the boundary theory: $\epsilon\sim 1/\delta$.   Before determining its precise value, let us first discuss  the third quantity $\mathcal N$.  
The total energy of a CFT state is through the operator-state correspondence determined by the scaling dimension $\Delta$ of the corresponding operator. The holographic dictionary relates the dimension $\Delta$ to the mass of the dual field in the bulk: for large scaling dimensions $\Delta\gg d$ the relationship is $\Delta \sim M L$.  Hence $\Delta$ counts the energy in terms of the IR cut-off scale $1/L$.  
The quantity $\mathcal N$ can be viewed as the UV analogue of $\Delta$: it counts the energy $E$ in terms of the UV cut-off $\epsilon$
\begin{equation} 
\label{defN}
E=\mathcal N  \epsilon \, .
\end{equation}
Here $E$ represents the energy of the microscopic theory  
The excitation number $\mathcal N$ is linearly related to the conformal dimension, where the linear coefficient is given  by the ratio of the IR and UV energy scales. 

By increasing the energy $E$ one starts to excite more degrees of freedom in the microscopic theory and eventually all UV degrees of freedom are excited on the holographic surface at radius $R$. This corresponds to the creation of a black hole of size $R$.  We will choose to normalize $\epsilon$ so that for a black hole we precisely have ${\mathcal N}={\mathcal C}$.   The asymptotic form of the AdS-Schwarzschild metric is given by (\ref{metric}) with blackening factor\footnote{Here we consider  large black holes with horizon size $R_h\gg L$, so that we can ignore the constant term in $f(R)$. We will include this term later in section \ref{sec:physicalimplications} and discuss its microscopic interpretation.}
\begin{equation}
\label{blackening}
f(R) = \frac{R^2}{L^2} - \frac{16 \pi G_d E }{(d-2) \Omega_{d-2} R^{d-3}} \, .
\end{equation}
 It is now easy to deduce the normalization of $\epsilon$ for which ${\mathcal N}= {\mathcal C}$ when $f(R)=0$.  We find that for super-AdS scales the excitation energy $\epsilon$ of the UV degrees of freedom equals
\begin{equation} \label{epsilonads}
\epsilon 
= (d-2)\frac{R}{L^2 } \  \qquad  \text{for}  \qquad R\gg L \, .
\end{equation}  
As we will discuss in section \ref{sec:geometric}, all the three quantities, $\mathcal C$, $\mathcal N$ and $\epsilon$ can be defined geometrically, where the latter two make use of an appropriately chosen reference metric.  Before discussing these geometric definitions, let us first consider the specific example of AdS$_3$/CFT$_2$, where our definitions will become more transparent. This case will also be  of crucial importance to   our study of holography in other spacetimes.

\subsection{An example: \texorpdfstring{AdS$_3$/CFT$_2$}{AdS(3)/CFT(2)}}
\label{sec:BTZ}
 
 To illustrate the meaning of the various quantities introduced in the previous section, let us consider the  AdS$_3$/CFT$_2$ correspondence. In particular we will further clarify the relation between $\mathcal C$ and $\mathcal N$, on the one hand, and the central charge $c$ and scaling dimensions $\Delta$ of the 2d CFT, on the other hand.  The metric for a static asymptotically  AdS$_3$ spacetime can be written as
  \begin{equation} \label{BTZmetric}
  ds^2 = \left ( \frac{r^2}{L^2} - \frac{\Delta - c/12}{c/12} \right) dt^2 + \left ( \frac{r^2}{L^2}  - \frac{\Delta - c/12}{c/12} \right)^{-1} dr^2 + r^2 d \phi^2 \, ,
  \end{equation}
    where $c$ and $\Delta$ are the  central charge and the scaling dimension in the dual 2d CFT.   $\Delta = 0$ corresponds to empty AdS; $\Delta \leq c/12$ is dual to a conical defect in AdS$_3$; and for $\Delta \ge c/12    $ the metric represents a BTZ black hole \cite{Banados:1992wn}. 
The holographic dictionary between AdS$_3$  and CFT$_2$ is well understood and states that  (see \cite{Kraus:2006wn} for a review)
 \begin{equation}
 \begin{aligned}
&\text{central charge:} \qquad    \quad  \,      \frac{c}{12} = \frac{2\pi L}{16 \pi G_3}  \,,\label{BrownHen}\\
&\text{scaling dimension:} \qquad      \Delta  -   \frac{c}{12 }  =  E L         \,.
 \end{aligned}
 \end{equation}
The first equation is just the Brown-Henneaux formula, and the second equation is the standard relation between the energy on the cylinder  and the scaling dimension $\Delta$.
The energy $E$ corresponds to the ADM energy of the bulk spacetime.   The holographic quantities $\mathcal C$ and $\mathcal N$ for AdS$_3$ are easily determined from  the   expressions    (\ref{nodof}) and~(\ref{defN}) 
\begin{equation}
 \begin{aligned}
&\text{number of d.o.f.:}    & \mathcal C &= \frac{2\pi r }{16 \pi G_3} = \frac{c}{12} \frac{r}{L}  \, ,  \\
&\text{excitation number:}    &\mathcal N &=  \,  E\, \frac{L^2}{r} \, \,   \, \,  =  \left (   \Delta - \frac{c}{12} \right) \frac{L}{r}   \,.
\label{excnumber1}
 \end{aligned}
 \end{equation}
 Note that excitation number can be negative, because $E$ is negative for empty AdS and the conical defect spacetime.
The reason for the increase in the number of degrees of freedom by $r/L$ is that the cut-off CFT at radius $r$ allows each field theoretic degree of freedom to have $r/L$ modes. Holographic renormalization (or the UV-IR connection) now tells us that for larger distances in the bulk the modes in the cut-off CFT carry a higher energy, given by $r/L^2$.  This means that at larger distances fewer UV degrees of freedom are excited for a state with fixed energy. Therefore, the excitation number at a radius $r>L$ decreases  by a factor  $L/r$ with respect to the value at the AdS radius. 

We see that $\mathcal C$ and $\mathcal N$  are rescalings of the central charge and the scaling dimension, respectively.   Since the rescaling is exactly opposite, the Cardy formula in 2d CFT remains invariant, and   can hence also be expressed in terms of $\mathcal C$ and $\mathcal N$ 
    \begin{equation}
  \label{CHR-formula}  
  S = 4 \pi \sqrt{\mathcal C \mathcal N} = 4 \pi \sqrt{\frac{c}{12} \left ( \Delta - \frac{c}{12} \right) } \, .
  \end{equation}
 To see that this  formula correctly reproduces the Bekenstein-Hawking entropy (\ref{BHformula}) \cite{Strominger:1997eq} one can use that  the following relations hold at the horizon of the BTZ black hole
\begin{equation}
   \Delta - \frac{c}{12} = \frac{r_h^2}{L^2} \frac{c}{12} \qquad   \text{hence} \qquad
\, .\mathcal N = \mathcal C  .
\end{equation}
Hence our definition of $\cal N$ indeed equals $\cal C$ on the horizon for this 3d situation.

\subsection{Geometric definition and generalization to sub-AdS scales}
\label{sec:geometric}

The number of degrees of freedom, UV cut-off energy and excitation number, as defined above, are general notions that in principle apply to any microscopic theory. A natural question is whether these concepts can be generalized to the microscopic theories on other holographic screens than those close to the AdS boundary. Our reason for introducing the quantities $\mathcal C$, $\mathcal N$ and $\mathcal \epsilon$ is that they can be defined in terms of the local geometry near the holographic surface $\mathcal S$.  In this subsection we will present this geometric definition and verify that it holds for large holographic screens.  Our next step is to postulate that the same geometric definition holds for other situations, in particular for the microscopic theory that lives on holographic screens at sub-AdS scales.

\begin{figure}
	\centering
	\includegraphics
		[width=.35\textwidth]
		{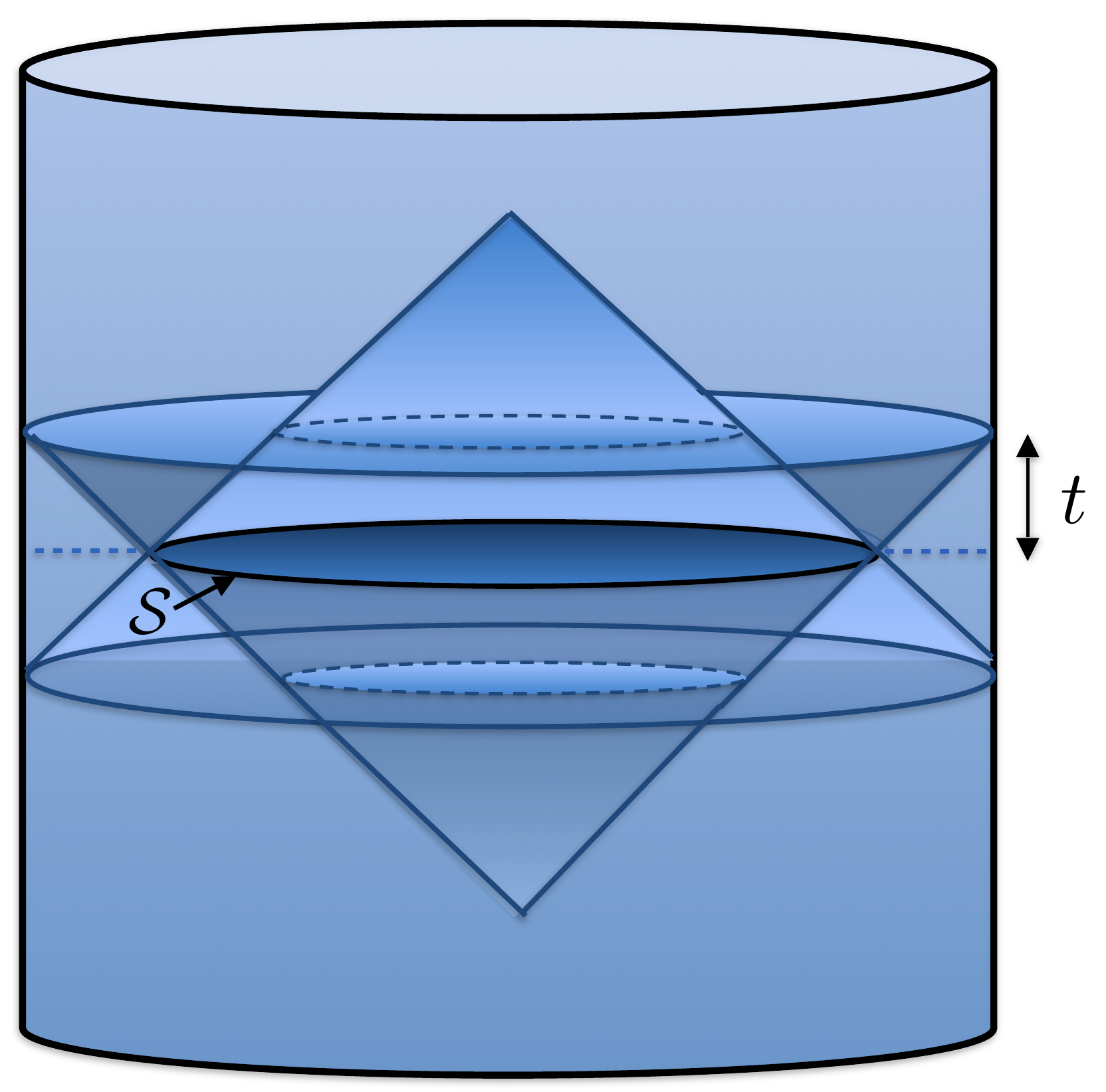}
\caption{\small \it A large causal diamond in AdS consisting of the domain of dependence of the ball  bounded by the holographic screen $\mathcal S$. The ball and the screen lie in the $t=0$ time slice, and are centered around the origin. The distance of $\mathcal S$ to the AdS boundary can be characterized by the time $t$ at which the outward future lightsheet reaches the AdS boundary.}
	\label{fig:largediamond}
\end{figure}

Our geometric definition makes use of the causal diamond that can be associated to the holographic screen. Causal diamonds play an important role in the literature on holography because of their 
invariant light-cone structure \cite{Bousso:2002ju,Bousso:1999xy,Banks:2011av,Banks:2013qpa}. Given a spherical holographic screen $\mathcal S$ with radius $R$ on a constant time slice of a static spherically symmetric spacetime,   the  associated causal diamond consists of the future and past domain of dependence of the ball-shaped region  contained within $\mathcal S$. For a large holographic screen in an asymptotically AdS spacetime, the corresponding causal diamond is depicted in Figure~\ref{fig:largediamond}. As shown in this figure, the distance to the AdS boundary can be parametrized by the time $t$ at which the extended lightsheets of the   diamond intersect the boundary.  The coordinate~$t$ corresponds to the global AdS time   and also gives a normalization of the local time coordinate near the screen $\mathcal S$. It is with respect to this time coordinate that we define the energy $E$. 

\begin{figure}
	\centering
	\includegraphics
		[width=.35\textwidth]
		{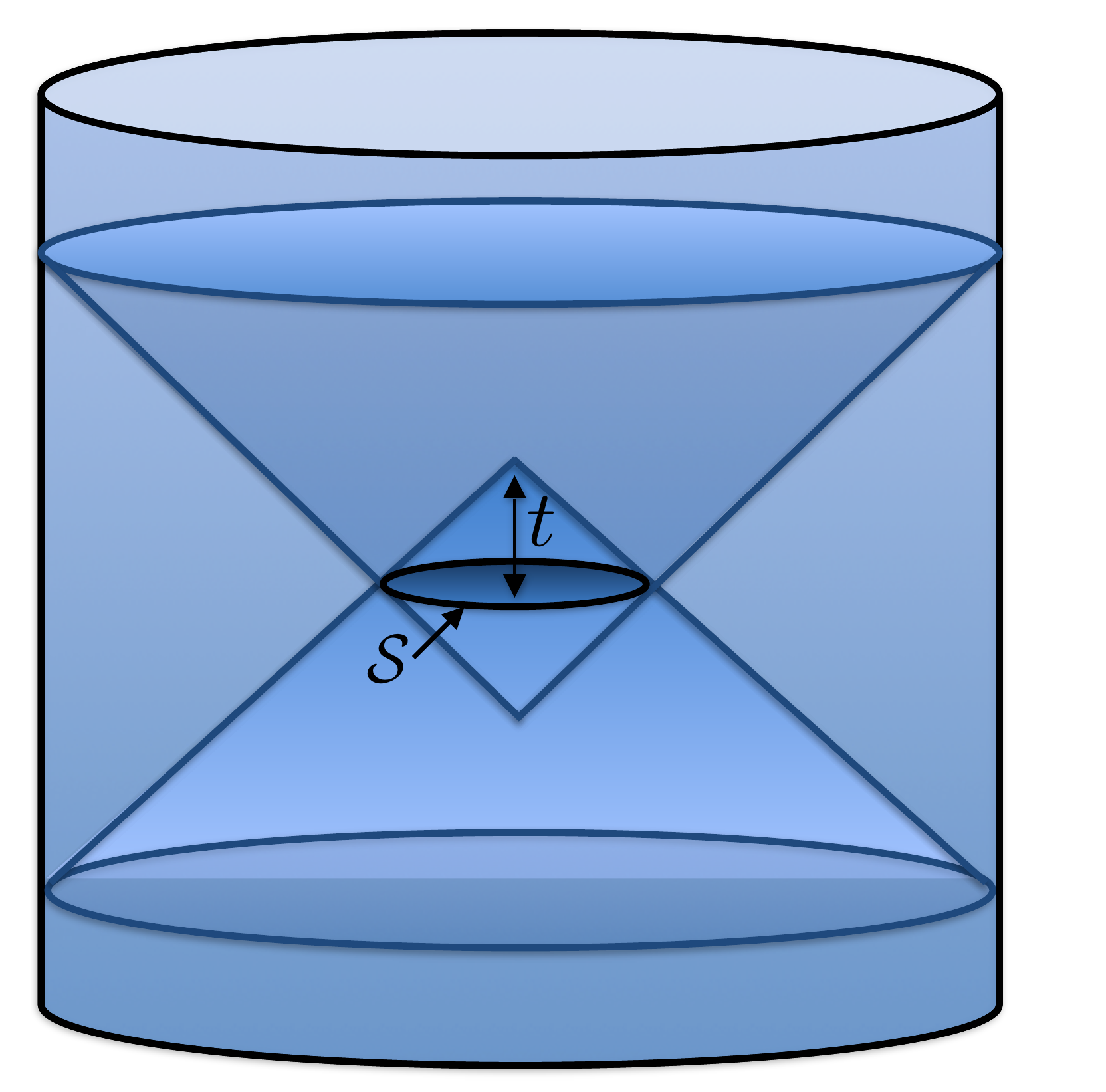}
\caption{\small \it A small causal diamond in AdS consisting of the domain of dependence of the ball  bounded by the holographic screen $\mathcal S$. The ball lies in the $t=0$ time slice, and is centered around the origin. The location of $\mathcal S$ with respect to the origin can be parametrized by the time $t$ at which the inward future lightsheet arrives at the origin.}
	\label{fig:smalldiamond}
\end{figure}

For   holographic screens at sub-AdS scales we will   introduce  a similar causal diamond. Except in this situation we define the time coordinate $t$ with respect to the local reference frame in the origin. For empty AdS this time coordinate is again given by the global AdS time $t$. The  causal diamond associated to a holographic screen at a small radius $R\ll L$ is depicted in Figure \ref{fig:smalldiamond}.

Let us consider the rate of change of the number of degrees of freedom $\mathcal C$ along a null geodesic  on the future horizon of the diamond. 
The time  $t$ can be used as a (non-affine) parameter along the null geodesic. For   metrics of the form (\ref{metric}) it is related to the radius $R$ by $dt = \pm dR /f(R)$, where the sign is determined by whether the null geodesic is outgoing (=plus) or ingoing (=minus). The rate of change with respect to $t$ can   thus be positive or negative depending on whether the time $t$ is measured with respect to  the origin or infinity, respectively. 
Our definitions for the excitation number $\mathcal N$ and excitation energy $\epsilon$ are chosen such that the following identity holds
\begin{equation}  \label{derivative2}
\left|\frac{d \mathcal C}{dt}\right|   =  \left (  \mathcal C - \mathcal N   \right) \epsilon   \,   ,
\end{equation}
where the absolute value is taken to ensure that $\epsilon$ is positive. This definition is motivated by the fact that the quantity $d\mathcal C/dt$ vanishes on the horizon of a black hole, if the horizon size is equal to $R$.  In this way it follows that on the horizon $\mathcal N = \mathcal C$.

The identity (\ref{derivative2}) is not yet sufficient   to fix the values of $\mathcal N$ and $\epsilon$. We need to specify for which geometry the excitation number $\mathcal N$ is taken to be zero. In other words, we need to introduce a reference metric that defines the state of zero energy. One could take this to be the empty AdS geometry. However, to simplify the equations  and clarify the discussions in the   subsequent sections
we will take a different choice for our reference geometry. For    super-AdS regions  the reference metric can be found by only keeping the leading term for large $R$ in the function $f(R)$. Whereas for   sub-AdS regions with $R\ll L$ we   take  the Minkowski metric to be the reference metric. Thus for these two cases the reference geometry has the form (\ref{metric}) where the function $f(R)$ is given by
\begin{equation}
\label{reference}
f_0(R) =\left\lbrace 
\begin{array}{c}    \!\!
\quad 1\ \, \qquad \text{for} \quad R\ll L\, ,\\
\!\! {R^2/L^2} \ \quad \text{for} \quad R\gg L\, . 
\end{array}\right.
\end{equation}
This geometry defines the state with vanishing   energy. We also take it to be the geometry where $\mathcal N$ is equal to zero. From (\ref{derivative2}) we thus conclude that the excitation energy is defined in terms of the reference metric via
\begin{equation}\label{excenergy}
 \epsilon = \left| \frac{1}{\mathcal C}\frac{d \mathcal C}{dt}  \right|_{0}      \, .
\end{equation}
We can now use the fact that in the reference geometry $dt = \pm dR /f_0(R)$ to compute the 
excitation energy explicitly
\begin{equation} \label{eq:newdefepsilon}
\epsilon = {f_0(R)\over \mathcal C}	{d\mathcal C\over dR} =\ 
\left\lbrace \begin{array}{c}  \!
 (d-2)/R\ \, \qquad \text{for} \quad R\ll L\, , \\
  \!\!(d-2){R/L^2} \ \quad \,\text{for} \quad R\gg L\, . \end{array}\right.
\end{equation}
For small causal diamonds the dependence on $R$ for the excitation energy is quite natural, because   the radius of the holographic screen is effectively the only scale there is.  
What is remarkable, though, is that the excitation energy increases when the size of the screen decreases. This is opposite to the situation at super-AdS scales, because in that case the excitation energy increases with the size of the screen.

Another way to arrive at this identification is to use the fact that the metric outside a mass distribution at sub-AdS ($R \ll L$) as well as   super-AdS scales ($R \gg L $) takes the form (\ref{metric}) where the blackening function is given by 
 \begin{equation}  \label{blackeningfactorfull}
 f(R) = f_0(R)    - \frac{ 16 \pi G_d E  }{(d-2) \Omega_{d-2} R^{d-3}} \,. 
 \end{equation} 
This equation allows us to verify     our geometric definition (\ref{derivative2}) of the excitation energy $\epsilon$ per degree of freedom and excitation number $\mathcal N$, and show that it is consistent with the identity (\ref{defN}) that expresses the total excitation energy as $E={\mathcal N}\epsilon$.  Using the fact that along a null geodesic $dt =\pm dR/f(R)$ one can derive the following relation\footnote{A similar observation was made by Brewin \cite{Brewin}, who noted that the ADM  mass is proportional to the rate of change of the area of a closed ($d-2$)-surface with respect to the geodesic distance. } 
\begin{equation} \label{derivative1}
\left| \frac{d \mathcal C}{dt} \right|   =   
\left| \frac{d \mathcal C}{dt} \right|_{0}  - E \, .
\end{equation}    
The first term on the right hand side is the contribution of the reference spacetime with blackening factor $f_0(R)$. By inserting the geometric definition (\ref{excenergy}) for $\epsilon$ and the definition  (\ref{defN}) of $\mathcal N$  into the equation above it is easy to check that this  reproduces the relation (\ref{derivative2}). In the following section we will provide further evidence for these relations for $\mathcal C$, $\mathcal N$ and $\epsilon$ by showing that the super-AdS and sub-AdS regions can be related through a conformal mapping that preserves the number of degrees of freedom $\cal C$ as well as the excitation number $\mathcal N$.

\section{Towards holography for non-AdS spacetimes}
\label{sec:towardsholographynonads}

We start this section by presenting two related conjectures that allow us to connect the physical properties of the microscopic theories that live on holographic screens in different spacetimes. In particular, we argue that the holographic theories for
 two spacetimes that are related by a Weyl transformation  have identical microscopic properties when the Weyl factor equals one on the corresponding holographic screens. We will apply this conjecture to obtain insights into the holographic theories for non-AdS spacetimes by relating them to the familiar case of AdS holography. 
We are especially interested in the holographic properties of sub-AdS regions, Minkowski  and  de Sitter space. We will describe a conformal mapping between AdS space (at super-AdS scale) on the one hand and AdS space (at sub-AdS scale), dS space or Minkowski space on  the other hand. We will use this mapping to derive a correspondence between the holographic descriptions of these spaces. Specifically, we will identify the quantities $\mathcal N$ and $\mathcal C$ in the two conformally related spacetimes.  

Other approaches towards non-AdS holography which also invoke holographic screens include the early work \cite{Bousso:1999cb} and more recently the Holographic Space Time framework \cite{Banks:2011av, Banks:2013qpa}.
A separate line of research has focused on the generalization of the Ryu-Takayanagi proposal in AdS/CFT \cite{Ryu:2006bv,Ryu:2006ef} to more general spacetimes. 
In particular, suitable holographic screens may be used to anchor the bulk extremal surfaces, whose areas are conjectured to provide a measure for the entanglement entropy of the holographic dual theory.  
See for example \cite{Li:2010dr,Shiba:2013jja,Miyaji:2015yva} for both bulk and boundary computations of this proposal.
In addition, the work of \cite{Sanches:2016sxy,Nomura:2016ikr,Nomura:2017fyh,Nomura:2018kji}  uses geometric aspects of carefully defined holographic screens to both establish and infer properties of the holographic entanglement entropy.
In the conclusion, we will point out how some of the conclusions of these works overlap with our own.

\subsection{A conjecture on  the microscopics  of conformally related spacetimes}
\label{sec:conjecture-on-micro}

In the previous section we showed that the number of holographic degrees of freedom $\mathcal C$ is given by the area of the holographic screen, and hence is purely defined in terms of the induced metric on $\mathcal S$. 
    Our definitions of $\mathcal N$ and $\epsilon$, on the other hand,  depend in addition on the geometry of the reference spacetime. Schematically  we have
\begin{equation}
\begin{aligned}
\mathcal C = \mathcal C \,  (  g , \mathcal S)  \,,  \qquad 
\epsilon = \epsilon \, (  g,   g_0 , \mathcal S) \, , \qquad 
\mathcal N = \mathcal N \,  (  g,   g_0, \mathcal S) \, , 
\end{aligned}
\end{equation}
where $g$ is the metric of the spacetime under consideration, and $g_0$ is the metric of the 
reference spacetime.  The presented definitions can in principle be applied to any static, spherically symmetric spacetime. The goal of this section is to gain insight into the nature and properties of these holographic degrees of freedom by comparing holographic screens in different spacetimes with the same local geometry. For this purpose it is important to note that the quantities $\mathcal C$ and $\mathcal N$ are purely expressed in terms of the (reference) metric on the holographic screen $\mathcal S$, without referring to any derivatives. One has for the case of spherically symmetric spacetimes
\begin{equation} \label{metricdef}
\mathcal C = \frac{1}{16 \pi G_d} \int_{\mathcal S} R^{d-2} \,  d\Omega_{d-2} \qquad \text{and} \qquad  \frac{ \mathcal N}{\mathcal C} = 1 - \frac{f(R)}{f_0(R)} \Big |_{\mathcal S}\, ,
\end{equation}
where $R$ is the radius of the spherical holographic screen. Here the second equation follows from   combining (\ref{derivative2}), (\ref{eq:newdefepsilon})  and   (\ref{derivative1}).  

The fact that $\mathcal C$ and $\mathcal N$ can be expressed purely in terms of the metric and not its derivative, suggests that  for holographic screens in two different spacetimes these quantities are the same if the local metrics on these holographic screens coincide. 

\begin{quote} \label{conjecture1}
\textbf{Conjecture I}: The holographic quantum systems on two holographic screens $\mathcal S$ and $\tilde{\mathcal S}$ in two different spacetime geometries $g$ and $\tilde g$, and with reference metrics $g_0$ and $\tilde{g}_0$, have the same number of (excited) degrees of freedom if the (reference) metrics are identical on the holographic screens~$\mathcal S$ and~$\tilde{\mathcal S}$:  
\begin{equation}
 g|_{\mathcal {S}}=\tilde{g}|_{\tilde{\mathcal {S}}},  \quad g_0|_{\mathcal{S}}=\tilde{g}_0|_{\tilde{\mathcal S}} \quad    \Rightarrow  \quad  {\mathcal C}  (g, \mathcal S) =  {\mathcal C} (\tilde g, \tilde{\mathcal S}), \quad {\mathcal N} (g,g_0, \mathcal S) = {\mathcal N} (\tilde g, \tilde g_0, \tilde{\mathcal S})   \, . \nonumber
\end{equation}
\end{quote}
The excitation energy $\epsilon$ will in general not be the same. In the specific cases discussed below, we find that $\epsilon$ is of the same order of magnitude in the two spaces, but in general differs by a (dimension dependent) constant  factor of order unity.  This motivates us to add to the conjecture that, in the specific cases we study, the excitation energies in the dual quantum systems differ only by an order unity constant: $   \epsilon (g,g_0, \mathcal S) \sim \epsilon (\tilde g, \tilde g_0, \tilde{\mathcal S})$.

Our goal is to study   the general properties of the microscopic theories that live on not just one, but a complete family of holographic screens in a given spacetime with metric $g$. By a complete family we mean that the holographic screens form a foliation of the entire spacetime. For a spherically symmetric spacetime one can choose these to be all spherical holographic screens centered around the origin. A particularly convenient way to find a mapping for all holographic screens is if the spacetime with metric $g$ is conformally related to $\tilde{g}$ via a Weyl rescaling\\[-2mm]
\begin{equation}
g=\Omega^2 \tilde{g}	 \, , 
\end{equation}
in such a way that the holographic screens $\mathcal S$ inside the spacetime geometry $g$ precisely correspond to the loci at which the conformal Weyl factor $\Omega$ takes a particular value. In other words, the holographic screens are the constant-$\Omega$ slices.  
 We denote this constant with $\Omega_{\mathcal S}$, so that $\mathcal S$ corresponds to the set of points for which $ \Omega=\Omega_{\mathcal S}$.  We thus have
 \begin{equation}
 \label{OmegaS}
  \left(\Omega - \Omega_{\cal S} \right)_{|{\mathcal S}} =0\\[-2mm]
 \end{equation}
 
 \begin{figure}[t] 
	\centering
	\includegraphics
		[width=0.7\textwidth]
		{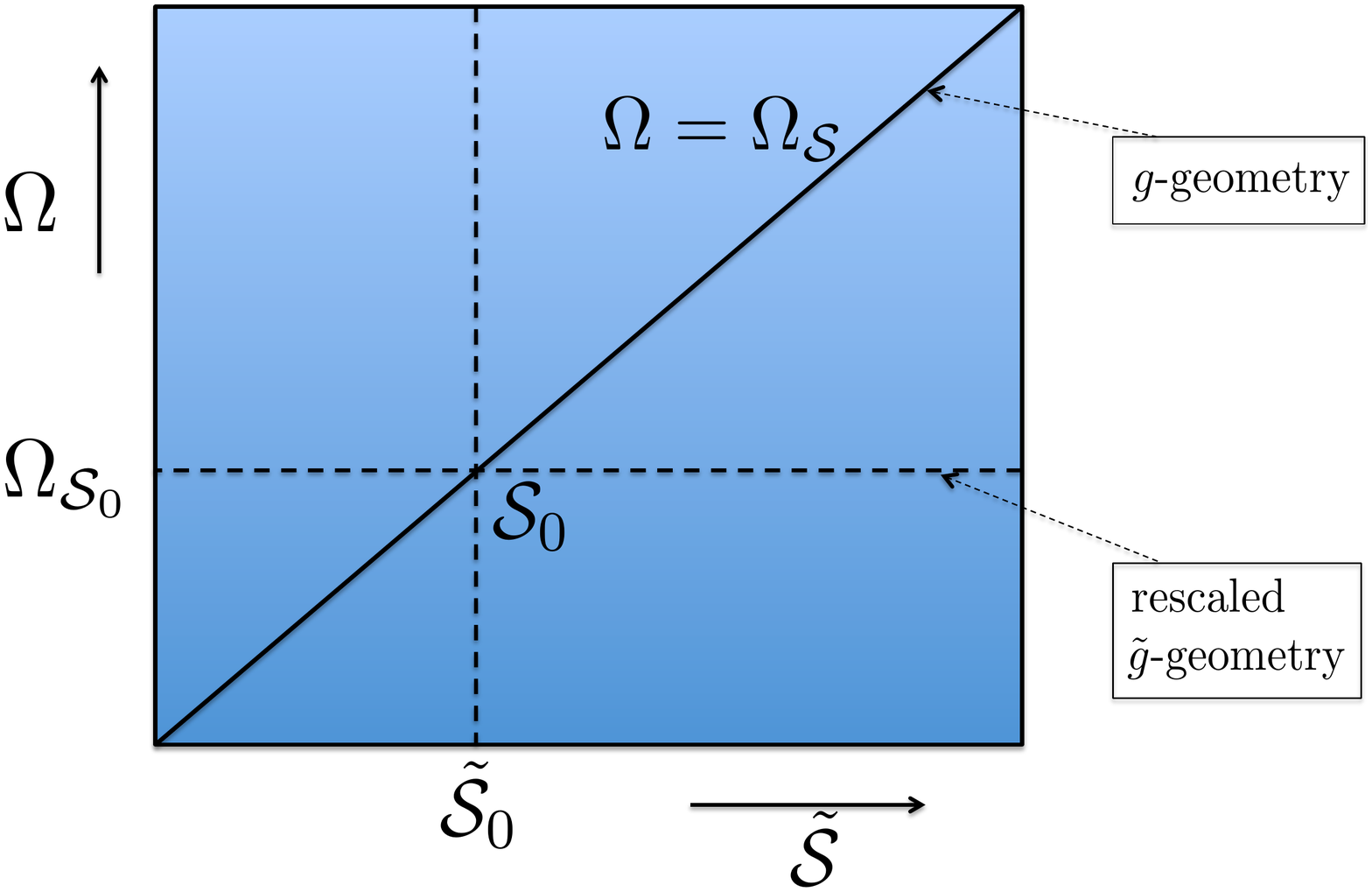}{}\\[-8mm]
	\caption{ \small \it{The family of rescalings of the geometry $\tilde g$ contains the geometry $g$ as a `diagonal' described by the equation $\Omega=\Omega_{\mathcal S}$. A horizontal line corresponds to a constant rescaling of $\tilde g$:  $\Omega_{\mathcal S_0} \tilde g$. On the holographic screen ${\mathcal S}_0$ the two geometries coincide: $g_{|\mathcal S_0} = \left(\Omega^2_{\mathcal S_0} \tilde g\right)_{|\mathcal S_0}$. By repeating this process for all values of $\Omega$ a foliation of the geometry $g$ is constructed and represented by the diagonal.
	}}
	\label{fig:family-of-geometries}
\end{figure}

\noindent The screens $\tilde{\mathcal S}$ inside the geometry $\tilde g$ are mapped onto a family of rescaled screens inside the rescaled geometries $\Omega^2 \tilde{g}$, where here $\Omega$ is taken to be constant on the $\tilde{g}$ geometry.  We will collectively denote the family of these screens by $\tilde{\mathcal S}$. The conformal map relates the screen $\mathcal S$ inside the geometry $g$ to one representative of this family, namely its image in the geometry $\Omega_{\mathcal S}^2 \tilde{g}$.  We will use the same notation for the holographic screen $\mathcal S$ and its image under the conformal mapping, and reserve the notation $\tilde{\mathcal S}$ for the family of screens inside the complete family of rescaled geometries $\Omega^2 \tilde{g}$.  
In fact, one can view $\tilde{\mathcal S}$ as the family (or equivalence class) of rescalings of the representative $\mathcal S$.  Together the set of all families $\tilde{\mathcal S}$ forms a foliation of the family of rescaled $\tilde{g}$-geometries, while the spacetime with metric $g$ is foliated by the particular representative $\mathcal S$.  In this way the latter spacetime can be viewed as a `diagonal' inside the family of rescaled geometries $\Omega^2 \tilde {g}$ defined by taking $\Omega =\Omega_{\mathcal S}$.  This situation is illustrated in Figure \ref{fig:family-of-geometries}.

An additional requirement on the conformal relation between the two spacetimes is that the reference metrics $g_0$ and $\tilde{g}_0$ are related by the conformal transformation\\[-2mm] \begin{equation}
	g_0=\Omega^2 \tilde{g}_0
\end{equation}
with the same Weyl factor $\Omega$. 
In this situation  each screen $\mathcal S$ in the spacetime with metric $g$ has a corresponding holographic screen   
in the spacetime with metric $\Omega^2_{\mathcal S} \tilde{g}$ satisfying the requirements of our first conjecture.  Namely, one has  
\begin{equation}
g _ {|\mathcal S} =\left(\Omega^2 \tilde{g}\right) \!_{|\mathcal S}	
= \left(\Omega^2_{\mathcal S} \, \tilde{g}\right) \! _{|\mathcal S}
\end{equation}
and a similar relation holds for the reference metrics.   
The rescaling with $\Omega_{\mathcal S}$ changes the area of the holographic screen in the $\tilde{g}$ geometry, and hence its number of degrees of freedom $\mathcal C$, so that it precisely matches the number on the screen $\mathcal S$ in the geometry $g$.  

We are now ready to state our second conjecture

\begin{quote} \label{conjecture2}
\textbf{Conjecture II}:  For  two conformally related spacetimes $g$ and $\tilde g$ with $g=\Omega^2 \tilde{g}$ the holographic quantum systems on a holographic screen $\mathcal S$ in the spacetime with metric $g$ and its image in the spacetime with metric $\Omega^2_{{\mathcal S}}\tilde{g}$ have the same number of (excited) degrees of freedom:   
\begin{equation}
g=\Omega^2 \tilde {g}  \quad \Rightarrow  \quad  \mathcal C  (g, \mathcal S) =  \mathcal C (\Omega^2_{\mathcal S} \, \tilde g, \mathcal S), \quad \mathcal N (g,g_0, \mathcal S) = \mathcal N (\Omega^2_{\mathcal S}\,\tilde g, \Omega^2_{\mathcal S}\tilde g_0, \mathcal S)   \, . \nonumber
\end{equation}
\end{quote}
In the rest of the paper we will postulate that conjecture II is true and apply it to gain insights into the microscopic features of the holographic theories for a number of ~spacetimes, including de Sitter and Minkowski space.  Since the metrics on the holographic screens $\mathcal S$ coincide, one can imagine cutting the two spacetimes with metric $g$ and $\Omega^2_{\mathcal S_0}\tilde{g}$ along the slice $\mathcal S_0$ and gluing them to each other along $\mathcal S$. In Figure \ref{fig:family-of-geometries} this amount to first going along the diagonal $\Omega =\Omega_{\mathcal S}$ and then continue horizontally along $\Omega =\Omega_{\mathcal S_0}$.  The resulting metric will be continuous but not differentiable. Hence, to turn this again into a solution of the Einstein equations, for instance, one should  add a mass density on $\mathcal S$, where we assume that the geometries $g$ and $\Omega^2_{\mathcal S} \tilde g$  are both solutions.

\subsection{An example: \texorpdfstring{$AdS_{d}\times  {S}^{p-2}  \cong AdS_{p}\times  {S}^{d-2}$}{AdS(d)xS(p-2)=AdS(p)xS(d-2)}} 
Let us illustrate the general discussion of the previous subsection with an example.  In  Appendix \ref{embedding} we discuss a class of spacetimes that are all  conformally related to (locally) AdS spacetimes. One particular case is the conformal equivalence 
\begin{equation}
  \textAdS_{d}\times S^{p-2}  \cong \textAdS_{p}\times S^{d-2}  \, . 
\end{equation}
The holographic screens $\mathcal S$ have the geometry of $S^{d-2}\times S^{p-2}$. Hence, the conformal map exchanges the spheres inside the  AdS spacetime with the sphere in the product factor. We identify the left geometry with 
$g$ and the right with $\tilde g$,  and denote the corresponding radii by $R$ and $\tilde{R}$. The AdS radius and the radius of the spheres are all assumed to be equal to $L$.  The Weyl factor $\Omega$ is a simple function of $R$ or $\tilde R$. It is easy to see that
\begin{equation}
\label{OmegaR}
\Omega = {R\over L} = {L\over \tilde{R}}	
\end{equation}
 since it maps the $(d\!-\!2)$-sphere of radius $L$  onto one of radius $R$ and the $(p\!-\!2)$-sphere with radius $\tilde R$ to one with radius $L$. This equation is the analogue of (\ref{OmegaS}).  Note that the conformal map identifies the holographic screens with radius $L$ inside both AdS-factors.
 
 We now come to an important observation: the conformal map relates sub-AdS scales on one side to super-AdS scales on the other side. Hence it reverses the UV and IR of the two spacetimes. 
 In particular holographic screens with $\tilde{R}\gg  L$ are mapped onto screens with $R\ll L$. This means that we can hope to learn more about the nature of the microscopic holographic theory for holographic screens in a sub-AdS geometry
by relating it to the microscopic theory on the corresponding screens in  the geometry with metric $\Omega^2 \tilde{g}$, which live  at super-AdS scales. 
First let us compare the values of the cut-off energies  $\epsilon$ and $\tilde \epsilon$. For the situation with $R\ll L$ and $\tilde{R}\gg L$ we found in (\ref{eq:newdefepsilon}) that
\begin{equation}\label{eq:epsilon-discont}
\epsilon = (d-2){1\over R} \qquad \mbox{and} \qquad \tilde \epsilon = (p-2){\tilde{R}\over L^2} \, , \qquad \mbox{hence}\qquad   {\tilde{\epsilon}\over \epsilon } = {p-2 \over d-2} \, .  
\end{equation}
We have thus verified that the energy cut-offs $\tilde{\epsilon}$ and $\epsilon$, before and after the conformal map are of the same order of magnitude, but differ by a dimension dependent factor. The inversion of the dependence on the radial coordinate is qualitatively explained by the fact that the conformal mapping reverses the UV and IR of the two AdS geometries. 

 Next let us compare the central charges of the two CFTs.  Since we imposed that the number of degrees of freedom of the microscopic theories are the same on corresponding holographic screens,  it follows that the central charges of the two sides must be the same at $R=\tilde{R} =L$. Namely, for this value of the radius the number of holographic degrees of freedom is equal to the central charge of the corresponding CFT.   Since we keep the value of the $d+p-2$ dimensional Newton's constant $G_{d+p-2}$ before and after the conformal map fixed, one can indeed verify that the central charges $c$ and $\tilde{c}$ agree when the AdS radius is the same on both sides:
 \begin{equation}
{c\over 12} = {\Omega_{d-2} L^{d-2}\over 16\pi G_d} =  {\Omega_{d-2}\Omega_{p-2} L^{d+p-4} \over 16\pi G_{d+p-2}} =  {\Omega_{p-2} L^{p-2} \over 16\pi G_p}  = {\tilde{c}(L)\over 12} . 
\end{equation}
Here $G_d$ denotes Newton's constant on AdS$_d$, while $G_p$ equals Newton's constant on AdS$_p$.  We indicated that the central charge $\tilde{c}$ is computed for the $R=\tilde{R}=L$ slice.   But how does the number of degrees of freedom change as we move to  say $R=R_0\ll L$?  This leads to a rescaling of the geometry on the right hand side with $\Omega_0 = R_0/L$, and hence it changes the curvature radius of the AdS$_p$ geometry and correspondingly the value of the central charge $\tilde{c}$. Note that the radius of the $S^{d-2}$ is rescaled as well and now equals $R_0$: this affects the relationship between the Newton constants $G_{d+p-2}$ and $G_p$.  The radius of the $S^p$ on the left hand side is unchanged, however, so the relation between $G_{d+p-2}$ and $G_d$ is still the same. In this way one finds that the central charge of the CFT corresponding to the rescaled $AdS_p\times S^{d-2}$ geometry becomes
\begin{equation}\label{eq:relation-c-tildec}
{\tilde{c} (R_0)\over 12} 	 = {\Omega_{p-2} R_0^{p-2} \over 16\pi G_p}   = {\Omega_{d-2}\Omega_{p-2} R_0^{d+p-4} \over 16\pi G_{d+p-2}} = {\Omega_{d-2} R_0^{d-2} \over 16\pi G_{d}} \left({R_0 \over L}\right)^{p-2}.  
\end{equation}
Thus the effective central charge of the CFT corresponding to the rescaled $\tilde{g}$ geometry depends on the radius $R_0$.  Note that in the left geometry $g$ we are at sub-AdS scales, while on the right $\tilde{R}_0= L^2/R_0\gg L$. This means that on the right we can use our knowledge of AdS/CFT to describe the microscopic holographic degrees of freedom.  Our conjecture II states that the general features of the microscopic theories on both sides agree. In this way we can learn about the microscopic theories at sub-AdS scales. 
 
As shown in Appendix \ref{embedding}, one can construct more general conformal equivalences that instead of AdS$_d$ contain Mink$_d$ or dS$_d$. All these geometries can, after taking the product with  $S^{p-2}$,  be conformally related to again a product manifold of a locally AdS$_p$ geometry with a $S^{d-2}$. 
The required conformal mappings may be obtained via an embedding formalism, as explained in detail in the Appendix.  

In the rest of the paper we will focus on the particular case $p=3$.  For this situation we have even more theoretical control, because of the AdS$_3$/CFT$_2$ connection.  Note that in this case the central charge $\tilde{c}(R_0)$ in \eqref{eq:relation-c-tildec}  grows as $R_0^{d-1}$ and hence as the volume. Another important reason for choosing $p=3$ is that the $(p\!-\!2)$-sphere becomes an $S^1$, whose size can be reduced by performing a $Z_N$ orbifold with large $N$, while keeping our knowledge about the microscopic degrees of freedom. The latter construction, as well as the significance of the volume law for the central charge, will be explained in section \ref{sec:longstring}.

\subsection{Towards holography for sub-AdS, Minkowski and de Sitter space}
\label{sec:three-ex}

In subsection 3.1 we explained how to foliate a spacetime metric $g$ in holographic screens by using a family of Weyl rescaled metrics $\tilde{g}$.   We will now apply this construction to the cases of sub-AdS$_d$, Mink$_d$ and dS$_d$ with a Kaluza-Klein circle,  by making use of the following conformal equivalences (see Appendix \ref{embedding})
\begin{align}\label{eq:global-ads/ds-weyl-equivalence-p=3}
\begin{split}
  \textAdS_{d}\times S^{1} &\, \cong \, \textAdS_{3}\times S^{d-2}\, ,\\
   \textMink_d  \times S^{1} &\,\cong \,  \textBTZ_{E=0} \times S^{d-2} \, ,  \\
 \textdS_{d}\times S^{1} &\, \cong \,   \textBTZ  \times S^{d-2} \, . 
\end{split}
\end{align} 
For all these examples, the Weyl factor $\Omega$  is given by (\ref{OmegaR}). In this and the following sections we will denote the coordinate radius of AdS$_3$ by $r$ instead of $\tilde R$, while we keep $R$  as the radius in the spaces on the left hand side of (\ref{eq:global-ads/ds-weyl-equivalence-p=3}). 
Earlier versions of the conformal map between $dS_{d}\times S^{1}$ and  $BTZ \times S^{d-2}$ appeared in  \cite{Anninos:2011af, Hubeny:2009rc}. 

The conformal equivalences can be verified easily using the explicit metrics. One can represent the metrics of the spacetimes on the right hand side of \eqref{eq:global-ads/ds-weyl-equivalence-p=3} as
\begin{equation}\label{confmap1}
d \tilde{s}^2 = -\left({r^{2}\over L^2} - \kappa\right) dt^2  +\left( {r^{2}\over L^2}  - \kappa  \right)^{-1} dr^{2} + r^{2} d\phi^{2} + L^2d\Omega_{d-2}^2 \, ,
\end{equation}
with  $\kappa = -1, 0$ or $+1$. Here   $\kappa = -1$ describes pure $AdS_3\times S^{d-2}$ in global coordinates;  $\kappa=0$ corresponds to the so-called massless BTZ black hole (for which $E=0$ and $r_h=0$); and $\kappa = +1$ represents the metric of a BTZ black hole with horizon radius $r_h = L$.

We now rescale the metric by a factor $L^2/r^2$ and subsequently perform the coordinate transformation $R=L^2/r$:
\begin{equation}  \label{conformafactor1}
d  s^2 = \Omega^2  d \tilde {s}^2 \qquad \qquad \Omega = \frac{R}{L} = \frac{L}{r} \, .
\end{equation}
This leads to the metrics
\begin{equation}\label{confmap2}
d {s}^2 = -\left( 1 -  \kappa {R^{2}\over L^2}\right) dt^2  +\left( 1 - \kappa {R^{2}\over L^2}\right)^{-1} dR^{2} + R^{2}d\Omega_{d-2}^2  + L^2d\phi^{2} \, , 
\end{equation}
with
\begin{equation} \label{kappa1}
\kappa =
\begin{dcases*}
 \ -1 &  for  $\ \textAdS_{d}\times S^{1} $\\
\ \ \ 0 & for $\ \textMink_d  \times S^{1}$ \\
\ +1  & for $\ \textdS_{d}\times S^{1}$  
\end{dcases*}
\, .
\end{equation}
An important property of these conformal equivalences is that the radius is inverted, i.e. $R= L^2/r$.  The inversion of the radius means, for example,  that asymptotic infinity  in AdS$_3$ is mapped to the origin in AdS$_d$, and vice versa.
Note also that the horizon of the BTZ black hole ($r=L$) is mapped onto the horizon of dS$_d$ space ($R=L$). 

On the AdS$_3$/BTZ side the different values of $\kappa$ correspond to different states in the dual two-dimensional CFT. One can read off the scaling dimensions  by comparing the metric (\ref{confmap1}) with the asymptotically AdS$_3$ metric (\ref{BTZmetric}): $\Delta =c/12 \left (  1 + \kappa \right)$. Using this, we rewrite the conformally rescaled metric  as
\begin{equation}  \label{eq:masterformula}
d  {s}^2=\frac{R^2}{L^2}\!\!\left[-\!\left(\frac{r^2}{L^2}\!-\!\frac{\Delta-c/12}{c/12}\right)  dt^2\!+\!\left(\frac{r^2}{L^2}\!-\!\frac{\Delta-c/12}{c/12}\right)^{-1}\!\!\! \!dr^2+r^2d\phi^2\!+\!L^2d\Omega^2_{d-2}\right]\!.
\end{equation}
This metric turns into (\ref{confmap2}) for the following values of the scaling dimension:
\begin{equation}
\begin{aligned}
&  \Delta = 0 \,\,\,\, : \quad &&  \text{anti-de Sitter space} \, ,        \\
&  \Delta  =   \frac{c}{12} : \quad   &&\text{Minkowski space} \, ,    \\
& \Delta =  \frac{c}{6} \,\,\,: \quad && \text{de Sitter space} \, . 
\end{aligned}
\end{equation}
The physical implications of these observations will be discussed further below. 

\begin{figure}[t]
	\centering
	\includegraphics
		[width=1\textwidth]
		{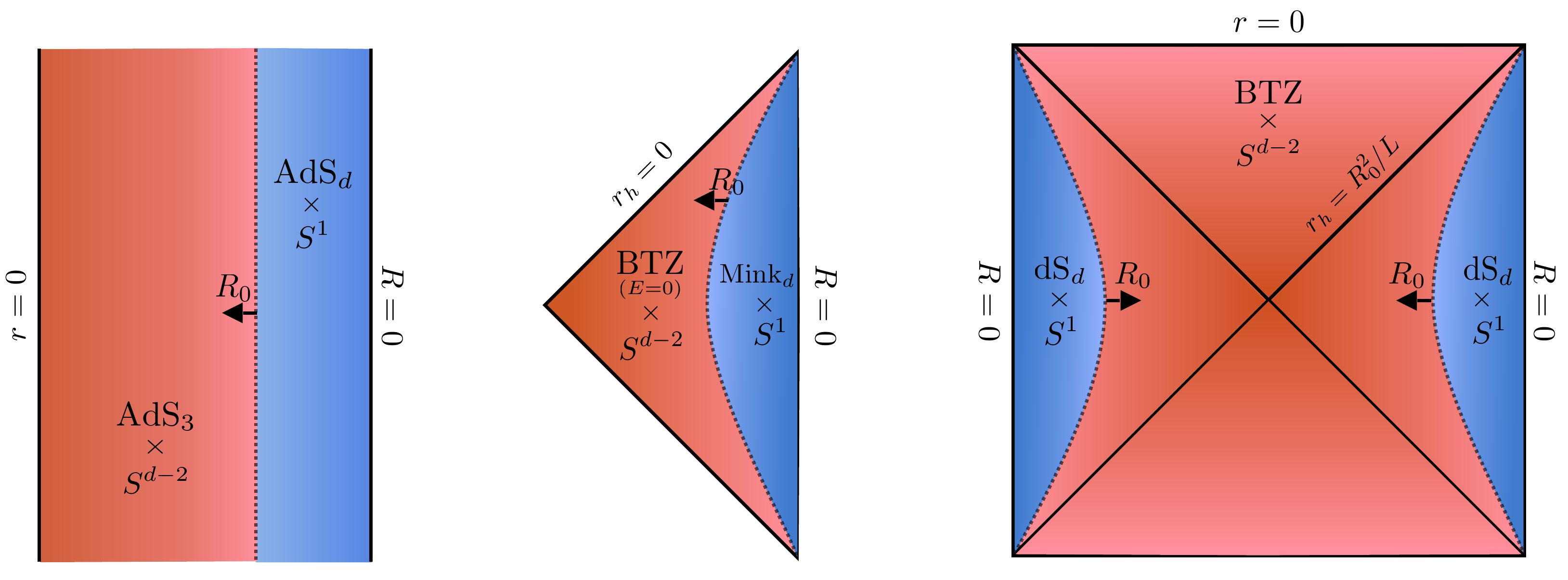}
	\caption{ \small \it{$\!\!\!$The first figure depicts, from left to right, the gluing of the $\left[0,r_0\right]$ region  of  Weyl rescaled $AdS_3\times S^{d-2}$ to the $[0,R_0]$ region of $AdS_d\!\times\! S^1$. The second figure illustrates, from left to right, the gluing of the $[0,r_0]$ region of massless (Weyl rescaled) $BTZ \times S^{d-2}$ to the $[0,R_0]$ region of $Mink_d \times S^1$. Finally, the third figure shows  the gluing of the $\left[0,r_0\right]$ region   of (Weyl rescaled) $BTZ\times S^{d-2}$ to the $[0,R_0]$ region of $dS_d\times S^1$. }}
	\label{fig:penrose-ads-ds-inside-out}
\end{figure}

To connect to our discussion in the previous section, consider our family of Weyl rescaled AdS$_3$ spacetimes obtained by taking a constant value of $R=R_0$. 
The AdS$_d$ slice at $R=R_0$ corresponds to the $r=r_0\equiv L^2/R_0$ slice in the Weyl rescaled AdS$_3$ geometry.
We now glue the region $0\leq r\leq r_0$ of the $AdS_3/BTZ\times S^{d-2}$ geometry to the $(A)dS_d/Mink_d\times S^1$ geometry, as illustrated in Figure \ref{fig:penrose-ads-ds-inside-out}.
We take $R_0<L$, so that the position at which the spacetimes are glued is at super-AdS$_3$/BTZ and sub-(A)dS$_d$ scales.
This allows us to interpret the microscopic description of sub-(A)dS$_d$ slices from a super-AdS$_3$ perspective through our conjecture. For this purpose we just need to give an interpretation to the metric \eqref{eq:masterformula} for $\ell\leq R_0\leq L$. This will be the main approach that we employ in section \ref{sec:longstring}.

However, already at this level we will make a couple of general remarks.
First of all, the metric within brackets is a locally AdS$_3$ spacetime. At super-AdS$_3$ scales we can thus use our knowledge of $\text{AdS}_3/\text{CFT}_2$ to interpret the geometry in terms of a microscopic theory. As discussed in section 3.2,  the rescaling of the metric with the scale factor $R_0^2/L^2$ effectively changes the curvature radius of the spacetime.  
In the microscopic quantum system this implies that the central charge of the $\text{CFT}_2$ now depends on the radius $R_0$ in the $d$-dimensional spacetime.  This $R_0$-dependence is given by the $p=3$ case of the general formula \eqref{eq:relation-c-tildec}. In fact, the central charge scales with the volume of the ball inside the holographic surface in AdS$_d$. As we will explain, the number of degrees of freedom  still grows like the area, as it should according to the holographic principle. 

Our goal in the remaining sections is to illuminate the nature of the holographic degrees of freedom that describe non-AdS spacetimes, by exploiting our conjecture and the conformal equivalence with the AdS$_3$ spacetimes. In particular we like to explain the origin of the reversal of the UV-IR correspondence at sub-(A)dS scales.  The usual concept of holographic renormalization takes us from the UV to the IR. In the sub-(A)dS case this means we have to start from a small radius and increase the radius as we go from UV to IR. The holographic principle then tells us that the number of degrees of freedom actually grows towards the IR. This is a rather unusual property compared with the familiar case of AdS/CFT, where as one moves from the UV to the IR one is integrating \emph{out} degrees of freedom.  In the case of sub-AdS or de Sitter holography, our conformal map suggests that we integrate \emph{in} degrees of freedom as we move from the UV to the IR  (a similar point of view was proposed  in \cite{Nomura:2017fyh,Anninos:2011af}). In the following section, we will describe a mechanism by which UV holographic degrees of freedom at the center of sub-AdS or de Sitter can be embedded in the IR holographic degrees of freedom at the (A)dS radius, thus still realizing the holographic principle.

\section{A long string  interpretation}
\label{sec:longstring}

In this section we will give a microscopic interpretation of the degrees of freedom of the holographic quantum system describing the metric (\ref{eq:masterformula}).
The identification of the correct degrees of freedom   enables us, through  our  conjecture II, to make some precise quantitative and qualitative statements about the holography for sub-AdS scales, Minkowski space and de Sitter space.
It will turn out that the microscopic quantum system underlying the metric in \eqref{eq:masterformula} has an interpretation in terms of so-called  `long strings'.
We will constructively arrive at these long strings as follows. We begin in the UV with a small number of degrees of freedom described by a seed CFT with a small central charge. As we go to larger distances we start taking symmetric products of this seed CFT. To arrive at the correct value of the number of degrees of freedom and excitation energy, we apply a so-called long string transformation. In this way we are able to build up the non-AdS spacetimes from small to large distances.  

In our construction we make use of three different length scales: a UV scale, an IR scale and an intermediate scale. These different scales are denoted by: \\[-4mm]
\begin{table}[H]
\centering
\begin{tabular}{lll}
 $\ell$: & \quad   \qquad UV scale & $=\quad $ short string length   \\ \\
$R_0$: & \quad    \qquad intermediate scale&  $=\quad $ fractional string length \\ \\
$L$: &   \quad   \qquad  IR scale & $=\quad $ long string length  \\
\end{tabular}
\label{tab:string-scales}
\end{table}
\noindent 
The terminology `short', `fractional' and `long' strings will be further explained below in our review of the long string phenomenon. We will take  $R_0$ and $L$ to be given by integer multiples of the short string length $\ell$ \begin{equation}
R_0=k\ell \qquad \text{and} \qquad L = N \ell  \, .
\end{equation}
The microscopic holographic theory can be described from different perspectives, and depends on which degrees of freedom one takes as fundamental: the short, the fractional or the long strings.  It turns out that the value of the central charge depends on which perspective one takes. The metric that was written in section \ref{sec:three-ex} will arise in the long string perspective. However, the short and fractional string perspective will turn out to give useful insights as well. In the following we will therefore use the string length as a subscript on the central charge to indicate in which perspective we are working.  For instance, the central charge in the fractional string perspective is written as $c_{R_0} (\cdot)$, where the value between brackets is a measure for the size of the symmetric product under consideration.  

As we will review in section \ref{sec:review-long-string},  the long string phenomenon relates the central charges and spectra of these differently sized strings according to
\begin{equation}
c_L (R_0)= \frac{k}{N} c_{R_0} (R_0)= \frac{1}{N} c_{\ell} (R_0) \, .
\end{equation}
 The relevance of the fractional string perspective consists of the fact that its central charge   always equals the total number of microscopic degrees of freedom $\mathcal C$ associated to a holographic screen at radius $R_0$. 
Nevertheless, we will argue below that the long string picture is more fundamental.
We will now start with explaining the long string phenomenon and these formulas in more detail, and we will also discuss aspects of the corresponding dual AdS$_3$ geometries.

\subsection{The long string phenomenon}\label{sec:review-long-string}
The long string phenomenon was originally discovered in \cite{Maldacena:1996ds}  and developed in detail in~\cite{Dijkgraaf:1996xw}.
The starting point is a so-called `seed CFT' with central charge $c_{\ell}$. 
Consider now the CFT that is constructed by taking a (large) symmetric product of the seed CFT, i.e.
$$
\textCFT^M/S_M \, .
$$
This symmetric product CFT has central charge $c_{\ell}(M)=Mc_{\ell}$.
Operators in this theory may now also have twisted boundary conditions in addition to ordinary periodic ones.
The resulting twisted sector, labeled by a conjugacy class of $S_M$, gives rise to long string CFTs.
The word `{long}' refers to the fact these sectors behave as if they were quantized on larger circles than the original seed CFT.
For instance, the twisted sector that corresponds to the conjugacy class consisting of $M$-cycles gives rise to a single long string that is $M$ times larger than the seed  (or short) string.
 As a result, the spectrum of modes becomes fractionated because the momenta are quantized on a circle of larger radius. Moreover, the central charge is reduced since the twisted boundary condition sews together  independent short degrees of freedom into  a single long degree of freedom.  
For consistency, the spectrum of the long string is subjected to a constraint
$$
P=L_0-\bar{L}_0=0\mod M  .
$$
This implies that the total momentum of a state on the long string should   be equal to the momentum of some state on the short string. 
However, due to its fractionated spectrum there are many more states on the long string for any given total momentum.
In fact,   in a large symmetric product CFT  the dominant contribution to the entropy of a certain macrostate comes from the longest string sector \cite{Maldacena:1996ds}.

In more detail, to project onto a long string sector of size $N$, one inserts degree $N$ twist operators in the symmetric product CFT
\begin{equation}\label{eq:twist-operator}
(\sigma_N)^{M/N}|0\rangle \, ,
\end{equation}
where $N$ is assumed to be a divisor of $M$. $\!$The twist operator $\sigma_N$ has conformal~dimension
\begin{equation}\label{eq:twist-op-dim-short}
\Delta_\ell=\frac{c_\ell(M)}{12}\left(1-\frac{1}{N^2}\right).
\end{equation}
The insertion of the twist operators has, as mentioned above, two important effects: it reduces the number of UV degrees of freedom and furthermore lowers their excitation energy. The reduction of the number of degrees of freedom is due to the fact that the spectrum becomes fractionated, which lowers the number of degrees of freedom by a factor $N$. The long string phenomenon thus operates as
\begin{align}\label{eq:long-string-pheno}
\begin{split}
\Delta_L-\frac{c_L(M)}{12}&=N\left(\Delta_{\ell}-\frac{c_\ell(M)}{12}\right) \, ,\\
c_L(M)&= \frac{1}{N}  \,  \, c_\ell(M)  \, .
\end{split}
\end{align}
Note that the conformal dimension and the central are rescaled in opposite direction. This implies in particular that the Cardy formula (\ref{CHR-formula}) is invariant under the long string transformation~\eqref{eq:long-string-pheno}. Since the long string central charge is smaller, the vacuum (or Casimir) energy in the CFT$_2$ is lifted to a less negative value.
Moreover, the   state  \eqref{eq:twist-operator} in the short string perspective coincides with the ground state in the long string CFT, as can  easily be verified by inserting \eqref{eq:twist-op-dim-short} into \eqref{eq:long-string-pheno} 
\begin{equation}
\Delta_L=N\left(\Delta_\ell-\frac{c_\ell(M)}{12}\right)+\frac{c_L(M)}{12}=0 \, .
\end{equation}
After having introduced the long and short string perspective, let us go to the intermediate or fractional string perspective. Instead of applying the long string transformation, one could  also consider an intermediate transformation, replacing $N$ in \eqref{eq:long-string-pheno} by $k\!<\!N$. This would only partially resolve the twist operator, which means that there still remains a non-zero conical deficit. 
The resulting fractional strings have size $R_0= kL/N$.
In this case, a twist operator will remain, whose conformal dimension is smaller than (\ref{eq:twist-op-dim-short}). Its presence indicates that the fractional string of length $R_0$ does not close onto itself and should be thought of as a fraction of a long string of length $L$.
The spectrum and central charge of the fractional string are related to those of the short string by
\begin{align}\label{eq:fractional-string-pheno}
\begin{split}
\Delta_{R_0}-\frac{c_{R_0}(M)}{12} & =k\left(\Delta_\ell-\frac{c_{\ell}(M)}{12}\right),\\
c_{R_0}(M)&=\frac{1}{k} \, \, c_{\ell}(M) \, .
\end{split}
\end{align}
The dimension of the remaining twist operator is obtained by inserting (\ref{eq:twist-op-dim-short})  into   the equation above
\begin{equation}\label{eq:twist-op-dim-fractional}
\Delta_{R_0}=\frac{c_{R_0}(M)}{12}\left(1-\frac{k^2}{N^2}\right) \, .
\end{equation}
We now turn to the AdS side of this story.
The  state   \eqref{eq:twist-operator} is   dual to a conical defect of order $N$ in AdS$_3$ \cite{Martinec:1998wm}.
The conical defect metric is simply given by the metric for empty AdS$_3$  with the following identification for the azimuthal angle
\begin{equation}
\phi\equiv \phi +2\pi/N \, .
\end{equation}
where $N = L /\ell$. For now we take $M=N$ so that $L$ becomes the size of the longest string.   Also, it plays the role of the AdS radius, since the symmetric product central charge is related to the AdS radius through the Brown-Henneaux formula:
$$
c_\ell(L)=\frac{2L}{3G_3} \, .
$$
Here we introduced a slightly different notation for the symmetric product central charge by replacing $M$ with the corresponding AdS radius. This notation will be used in the rest of the paper.

We can rewrite the AdS$_3$ metric with conical defect in the following way 
\begin{align}\label{eq:metric-long-string-from-seed-no-sphere}
\begin{split}  
ds^2& =-\left ({r^{\,2}\over L^2}+1\right) dt^2  +\left ({r^{\,2}\over L^2}+1\right)^{-1} \! \!dr^{2} +r^{2} d\phi^{2} \\
&=N^2\left[-\left ({\hat{r}^{\,2}\over \ell^2}+\frac{1}{N^2}\right) dt^2  +\left ({\hat{r}^{\,2}\over \ell^2}+\frac{1}{N^2}\right)^{-1} \! \!d\hat{r}^{2} +\hat{r}^{2} d\hat{\phi}^{2} \right],
\end{split}
\end{align}
where $\hat \phi \equiv \hat\phi +2\pi$, and
\begin{align}\label{eq:coord-transf-ell-to-L-no-sphere}
\begin{split}
\hat{r}&= r/N^2  \qquad  \text{and} \qquad \hat{\phi}=N\phi \, .
\end{split}
\end{align}
This rewriting illustrates the geometric analog  of taking an $N^{\text{th}}$ symmetric product and projecting to a long string sector of size $N$.
Indeed, the metric with curvature radius $\ell$ can be interpreted as the dual of the seed CFT. 
The multiplication by $N^2$ scales up the curvature radius to $L$, which is the geometric analog of the symmetric product. 
Moreover, the $1/N^2$ term in the $g_{tt}$ and $g_{\hat{r}\hat{r}}$ components is the analog of the insertion of the twist operator in the short string perspective.
Finally, the coordinate transformation leads us to the first metric in (\ref{eq:metric-long-string-from-seed-no-sphere}) whose covering space is dual to the long string CFT~\cite{Balasubramanian:2014sra}.

The analog of \eqref{eq:metric-long-string-from-seed-no-sphere} and \eqref{eq:coord-transf-ell-to-L-no-sphere} for a fractional string transformation is given by
\begin{align}\label{eq:metric-fract-string-from-seed-no-sphere}
\begin{split}
ds^2   &=  \frac{N^2}{k^2} \left[ -\left ({\tilde r^{\,2}\over R_0^2}+\frac{k^2}{N^2}\right) dt^2  +\left ({\tilde r^{\,2}\over R_0^2}+\frac{k^2}{N^2}\right)^{-1} \! \!d\tilde r^{2} +\tilde r^{2} d\tilde\phi^{2}\right]     \\
&=N^2 \left[-\left ({\hat{r}^{\,2}\over \ell^2}+\frac{1}{N^2}\right) dt^2  +\left ({\hat{r}^{\,2}\over \ell^2}+\frac{1}{N^2}\right)^{-1} \! \!d\hat{r}^{2} +\hat{r}^{2} d\hat{\phi}^{2}  \right],
\end{split}
\end{align}
where $\tilde{\phi}=\tilde{\phi}+2\pi /k$, and 
\begin{align}\label{eq:coord-transf-ell-to-R0-no-sphere}
\begin{split}
\hat{r}&= \tilde r/k^2 \qquad \text{and}  \qquad \hat{\phi}=k\tilde{\phi} \, .
\end{split}
\end{align}
The presence of the $k^2/N^2$ is the geometric manifestation of the fact that we have not fully resolved the twist operator.
In particular, it indicates that the fractional string does not close onto itself.
Since $\tilde{\phi}$ is periodic with $2\pi/k$,   a fractional string precisely fits on the conformal boundary $\tilde{r}=R_0$ in the covering space. 

 In the next section, we will also be interested in changing the size of the symmetric product. In particular, instead of multiplying the metric by $N^2$ we will also consider multiplication by $k^2$.  In that case, the conformal  factor in front of the first metric in \eqref{eq:metric-fract-string-from-seed-no-sphere}  is one. This describes the AdS dual of a single fractional string. 
This metric and its dual CFT interpretation will play an important role in our discussion of sub-(A)dS$_d$.

\subsection{Sub-AdS scales}\label{sec:sub-ads}
In this section we will put our conjecture II in section \ref{sec:conjecture-on-micro} to use.
By employing the long string mechanism explained in the previous section  we will give an interpretation of the holographic degrees of freedom relevant for sub-AdS scales.

We are interested in the slice $R=R_0 \le L$  in the $AdS_d\times S^1$ metric  
\begin{equation}\label{eq:metric-adsd-cirlce}
ds^2=-\left (1+{R^{\,2}\over L^2}\right) dt^2  +\left (1+{R^{\,2}\over L^2}\right)^{-1} \! \!dR^{2} +R^{2}d\Omega_{d-2}^2+ \ell^2 d\Phi^{2} \, ,
\end{equation}
with $\Phi\equiv \Phi +2\pi$.
In contrast to the metric (\ref{confmap2}),  here we have   adjusted the size of the transverse circle to $\ell\ll L$ in order to compactify to AdS$_d$ even at sub-AdS scales.
In the AdS$_3$ spacetime this can be achieved by the insertion of a conical defect, as in the first metric of (\ref{eq:metric-long-string-from-seed-no-sphere}). Combining the conformal map of section \ref{sec:three-ex} with equation (\ref{eq:metric-long-string-from-seed-no-sphere}), we can rewrite the metric  above as  
\begin{align}\label{eq:metric-long-string-from-seed}
\begin{split}
ds^2 
&=\frac{R^2}{\ell^2}\left[-\left ({\hat{r}^{\,2}\over \ell^2}+\frac{1}{N^2}\right) dt^2  +\left ({\hat{r}^{\,2}\over \ell^2}+\frac{1}{N^2}\right)^{-1} \! \!d\hat{r}^{2} +\hat{r}^{2} d\hat{\phi}^{2} + \ell^2d\Omega^2_{d-2}\right],
\end{split}
\end{align}
with $\hat{\phi}\equiv \hat{\phi}+2\pi$. Note that we have made the following identification between the radial and angular coordinates
\begin{equation}
R   = \frac{\ell^2}{\hat r} \,  \qquad \text{and} \qquad \Phi = \hat \phi \, . 
\end{equation}
For any fixed $R=R_0$  we obtain an equivalence between slices in $AdS_d\times S^1$ and in a conformally rescaled $AdS_3\times S^{d-2}$ spacetime with conical defect.
We will first consider the case $R=L$ to provide a CFT$_2$ perspective on the holographic degrees of freedom at the AdS$_d$ scale, and thereafter consider sub-AdS$_d$ scales.

Without the conformal factor and the conical defect, the $AdS_3 \times S^{d-2}$ metric with curvature radius $\ell$ is dual to the ground state of a seed CFT with central charge
\begin{equation}\label{eq:seed-central-charge}
{c_\ell(\ell)\over 12} = {2\pi \ell\over 16\pi G_3}= { A(\ell) \over 16\pi G_d} \, ,
\end{equation}
where we used  $1/G_3 = A(\ell)/ G_d$ and  $1/G_d =2\pi \ell / G_{d+1} $.
We imagine $c_{\ell}(\ell)$ to be a relatively small central charge, just large enough to be able to speak of a `dual geometry'.
Due to the presence of the transversal sphere, the multiplication of the seed metric by $N^2$ now scales up the central charge of the seed CFT by a factor $N^{d-1}$ if we choose to keep $G_{d}$ fixed:
\begin{equation}\label{eq:sym-prod-central-charge}
{c_\ell(L)\over 12} = N^{d-1}{c_\ell(\ell)\over 12}  \, . 
\end{equation}
As  explained in section \ref{sec:review-long-string}, we may interpret this rescaling as taking an $N^{d-1}$-fold  symmetric product of the seed CFT.
Additionally, the $1/N^2$ term in the metric (\ref{eq:metric-long-string-from-seed}) signals the presence of a twist operator in the dual CFT, that puts the system in a long string sector of the symmetric product CFT.
This reduces the central charge of the system and fractionates the spectrum of the theory, as expressed in \eqref{eq:long-string-pheno}.

Our conjecture relates the holographic quantities in the microscopic dual  of AdS$_3$ to   those in the dual of AdS$_d$.
There are two apparent problems when we think of the short string degrees of freedom in the symmetric product CFT as relevant to AdS$_d$. 
First, the symmetric product central charge \eqref{eq:sym-prod-central-charge} expresses a volume law for the number of holographic degrees of freedom at $R=L$. This number should however be related to the central charge of the CFT$_{d-1}$  which obeys an area law.
Moreover, the excitation energy required to excite the degrees of freedom at $R=L$ should be of the order $1/L$, but the seed degrees of freedom have an excitation energy of the order~$\epsilon_{\ell}\sim 1/\ell$.

The long string phenomenon precisely resolves both of these problems.
First, it reduces the volume law for the central charge to an area law
\begin{equation}
\frac{c_L (L)}{12} =  {1\over N} {c_\ell(L)\over 12}= \frac{A(L)}{16 \pi G_d} \, .
\end{equation}
Simultaneously, the long string phenomenon  give rises to a reduced excitation energy 
\begin{equation}\label{eq:excitation-energy-long-string}
\epsilon_L={1\over N} \epsilon_\ell = {d-2\over L} \, .
\end{equation}
The factor $(d-2)$ arises in AdS$_d$, as discussed around \eqref{eq:epsilon-discont}.
Concluding, the quantum system dual to the metric \eqref{eq:metric-long-string-from-seed} at $R=r=L$ has $c_L(L)/12$ holographic long string degrees of freedom, which may be excited with the lowest possible energy~$\epsilon_L$.
This  is consistent with our expectations for the dual quantum system of AdS$_d$ at $R=L$.
Since this discussion only concerns the number of holographic degrees of freedom and their excitation energies, our conjecture allows us to give a CFT$_2$  interpretation of the AdS$_d$ holographic degrees of freedom. \\

\noindent Next, to gain access to sub-AdS$_d$ scales, we will take $R=R_0<L$.
In this case, the seed metric is multiplied by $k^2$.
Analogously to the discussion above, this is interpreted as taking a $k^{d-1}$-fold symmetric product of the seed CFT.
The central charge of the symmetric product CFT is  
\begin{equation}  \label{eq:centralcharge-shortstring}
\frac{c_{\ell}(R_0)}{12}=k^{d-1}{c_\ell(\ell)\over 12} \, . 
\end{equation}
Performing the   coordinate transformation   \eqref{eq:coord-transf-ell-to-R0-no-sphere} on the metric  \eqref{eq:metric-long-string-from-seed}, we obtain the fractional string metric:
\begin{align}\label{eq:metric-fractional-string}
\begin{split}
ds^2  &= \frac{R^2}{R_0^2} \left [  -\left ({\tilde r^{\,2}\over R_0^2}+\frac{k^2}{N^2}\right) dt^2  +\left ({\tilde r^{\,2}\over R_0^2}+\frac{k^2}{N^2}\right)^{-1} \! \!d\tilde r^{2} +\tilde r^{2} d\tilde\phi^{2} + R_0^2d\Omega^2_{d-2} \right] \,  .  \\
\end{split}
\end{align}
The metric with curvature radius $R_0$ is dual to   fractional strings with central charge $c_{R_0}(R_0)$.
Fractional strings provide a  useful perspective on sub-(A)dS scales, since they represent the degrees of freedom that are directly related to the holographic quantities defined in section \ref{sec:generalfeatures}.
For instance, using \eqref{eq:fractional-string-pheno} one quickly verifies that 
\begin{equation} \label{nodofsubads}
\mathcal C=\frac{c_{R_0}(R_0)}{12}  .
\end{equation}
This shows that the fractional strings can be thought of as the sub-AdS analog of the super-AdS holographic degrees of freedom, as discussed by Susskind and Witten.
Indeed, fractional strings have a larger excitation energy than long strings:
\begin{equation}
\epsilon_{R_0} = \frac{d-2}{R_0}\,. 
\end{equation}
This is the same excitation energy as defined in (\ref{eq:newdefepsilon}) for sub-AdS scales. If we interpret $\epsilon_{R_0}$ as the UV cut-off at sub-AdS scales, then the fractional strings are the corresponding UV degrees of freedom. 
Further,    the fractional string quantities are related to the number of excitations by
\begin{equation}
 \mathcal N= \Delta_{R_0} - \frac{c_{R_0}(R_0)}{12} \, . 
\end{equation}
Similarly to \cite{Susskind:1998dq} a thermal bath of these fractional strings at temperature $\epsilon_{R_0}$ creates a black hole, and $\mathcal{N}=\mathcal{C}$     translates then to fractional string quantities as $\Delta_{R_0} = c_{R_0}/6.$ We will come back to this in more detail in section \ref{sec:bhentropy}.

Finally, we will discuss the long string perspective on sub-AdS holography.
In the long string picture the spectrum and central charge of the dual CFT are   given by
\begin{align}\label{cLR}
\begin{split}
 \Delta_L-\frac{c_L (R_0)}{12}  & = \:\frac{N}{k} \left(\Delta_{R_0}-\frac{c_{R_0}(R_0)}{12} \right)  \, ,   \\ 
c_L(R_0)& = \frac{k}{N} \, \, c_{R_0}(R_0)   \, .  
\end{split}
\end{align}
The factor $k/N$ in $c_L(R_0)$ expresses the fact that the fractional strings only carry a fraction of the long string central charge. 
Comparing with $c_L(L)$  we see that the number of long strings at $R=R_0$ is reduced from $N^{d-2}$ to $k^{d-2}$, of which only the fraction $k/N$ is accessible.

\begin{figure}[t]
	\centering
	\includegraphics
		[width=0.4\textwidth]
		{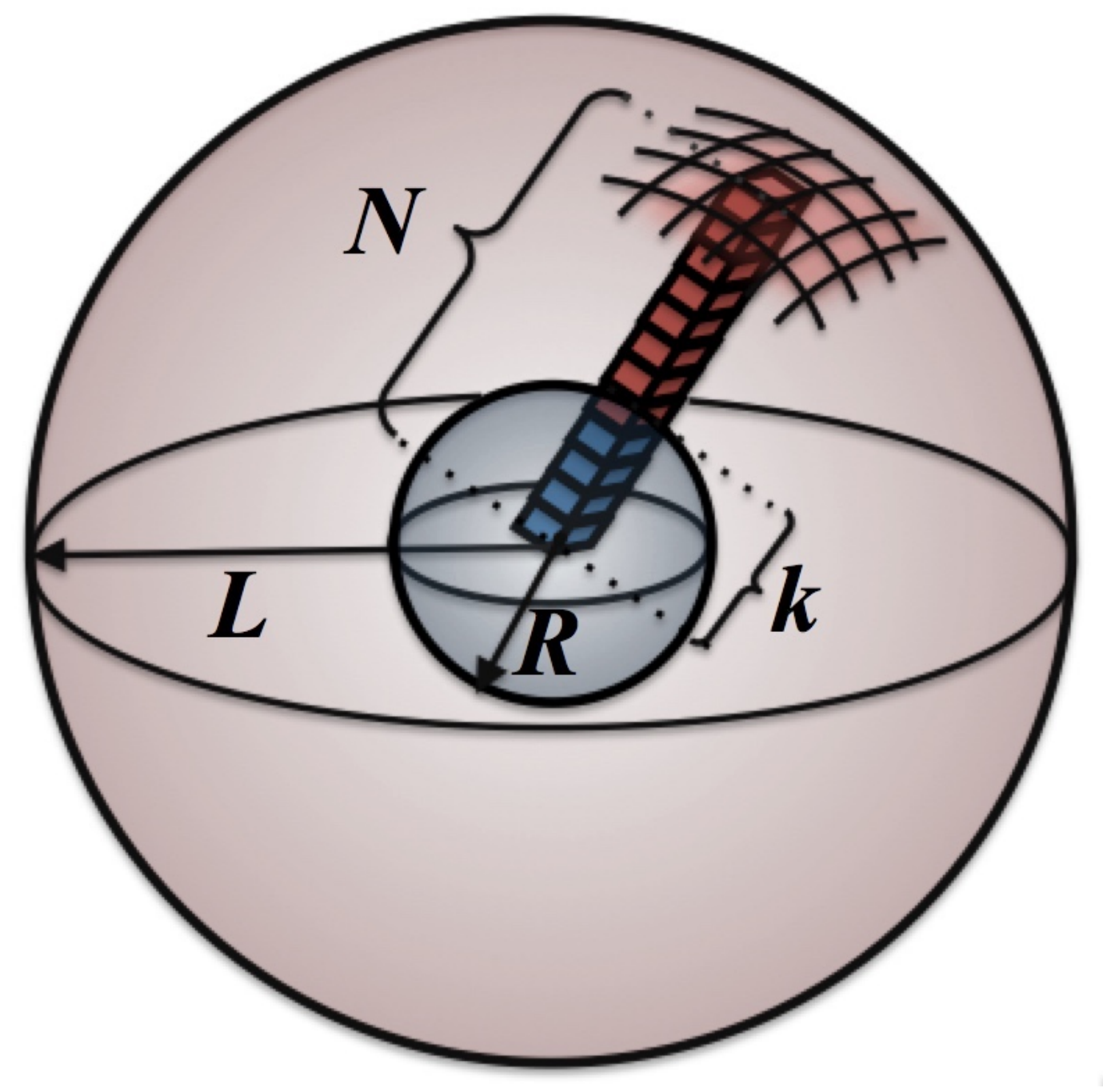}
	\caption{ \small \it The degrees of freedom on the holographic screen at radius $L$ consist of single `long strings'. The degrees of freedom on holographic screens at radius $R_0= k\ell$ with $k<N$ consist of `fractional strings'. The total number of long (or fractional) strings is proportional to the area of the holographic screen at radius $L$ (or $R_0$). }
	\label{fig:long-strings}
\end{figure}
 
Geometrically one can arrive at the long string perspective by applying the coordinate transformation \eqref{eq:coord-transf-ell-to-L-no-sphere} on the short string metric \eqref{eq:metric-long-string-from-seed}.
In this long string metric, the super-AdS$_3$ radial slice $r=L^2/R_0$ corresponds to the sub-AdS$_d$ slice $R=R_0$. 
 From the perspective of the long string we may therefore excite more modes.  In particular, the lowest energy excitation on the fractional string corresponds to the higher excited state on the long string:
\begin{equation}
\epsilon_{R_0} = \frac{N}{k}\epsilon_L. 
\end{equation}
In this sense, one can think of the excitations on fractional strings as bound states of the lowest energy excitations on the long strings.
This also explains that in the long string perspective the total number of degrees of freedom is still given by an area law, as opposed to $c_L(R_0)$, since we may excite more modes per each of the $c_L(R_0)$ degrees of freedom.

The explicit realization of the microscopic quantum system helps in understanding  the reversal of the UV-IR correspondence  at sub-(A)dS scales. 
The process of taking symmetric products and going to a long string sector achieves to integrate in degrees of freedom while moving to the IR.
More precisely, the symmetric product is responsible for integrating in degrees of freedom, where the order of the symmetric product determines the precise amount.
The (partial) long string phenomenon then reduces the excitation energy, realizing the step towards the IR.\footnote{Note that the larger the symmetric product becomes, the further one may move to the IR.} 
Moreover, from the AdS$_d$ perspective, it ensures that the number of degrees of freedom is given by an area law, and hence is in accordance with the holographic principle. 

We conclude this section by giving a master formula for the Weyl rescaled geometries we consider in this paper, which is a more refined version of \eqref{eq:masterformula}:
\begin{equation}\label{eq:metric-gen-conf-dim}
ds^2 =\frac{R^2}{L^2} \!\left[-\left(\frac{r^2}{L^2}-\frac{\Delta_{L}-\frac{c_L(R)}{12}}{\frac{c_L(R)}{12}}\right) \!dt^2+\left(\frac{r^2}{L^2}-\frac{\Delta_{L}-\frac{c_L(R)}{12}}{\frac{c_L(R)}{12}}\right)^{-1} \! \! \! \!dr^2+r^2d\phi^2+L^2d\Omega^2_{d-2}\right] \!.
\end{equation}
Here, $\phi\equiv \phi+2\pi/N$ as usual and $c_L(R)$ can be obtained by combining \eqref{nodofsubads} and \eqref{cLR}:
\begin{equation}\label{eq:cLR}
\frac{c_{L}(R)}{12} ={\Omega_{d-2} R^{d-1} \over 16\pi G_{d} L}={A(R) \over 16\pi G_{d} }{R \over L} \, .
\end{equation}
For sub-AdS, this metric is the long string version of \eqref{eq:metric-long-string-from-seed} and $\Delta_L=0$ accordingly.
However, this formula also captures all the non-AdS geometries we will consider in subsequent sections, and in addition provides them with a precise microscopic meaning.

\subsection{Minkowski space}
In this section, we briefly discuss the conformal map  between massless BTZ and Mink$_d$.
It allows us to phrase the holographic degrees of freedom relevant for Mink$_d$ in terms of  CFT$_2$ language introduced in the previous section.
Since we have discussed at length the conformal map between AdS$_3$ in the presence of a conical defect and sub-AdS$_d$, it is more convenient to think of the massless BTZ metric as the $N\to \infty$ limit of the conical defect metric.
We understand this limit as the $L\to \infty$ limit, keeping $\ell$ fixed, which on the AdS$_d$ side is of course the limit that leads to Minkowski space. 

Explicitly, the $\mathbb{R}^{1,d-1}\times S^1$ metric  
 \begin{align}  
\begin{split}
 ds^2=- dt^2  +d R^{2}+ R^2d\Omega^2_{d-2} +\ell^2 d\Phi^{2}  
\end{split}
\end{align}
is   equivalent to the $N\to \infty$ limit of the fractional string metric \eqref{eq:metric-fractional-string}, i.e.
 \begin{align}  
\begin{split}
 ds^2=\frac{R^2}{R^2_0}\left[- {\tilde r^{\,2}\over R_0^2} \,  dt^2  + { R_0^2 \over    \tilde r^{\,2}  }  \,d\tilde r^{2} +\tilde r^{2} d\tilde\phi^{2} + R_0^2d\Omega^2_{d-2}\right] \, .
\end{split}
\end{align}
In particular, at $R=\tilde{r}=R_0$ the fractional string metric is equivalent to the $\mathbb{R}^{1,d-1}\times S^1$ metric.

Our discussion on sub-AdS has taught us that the fundamental degrees of freedom are long strings of size $L$, of which we can only access a fraction at scales $R_0<L$.
In the case of Minkowski space, the long strings have an infinite length.
The infinite twist operator state in the short (or fractional) string perspective corresponds to the ground state of these infinitely long strings.
Therefore, in CFT$_2$ language Minkowski space can be thought of as the groundstate on infinitely long strings, which also has a vanishing vacuum (or Casimir) energy.
These two aspects are reflected in the equations:
$$
\Delta_{\infty}=c_\infty(R_0)=0.
$$
The vanishing of $c_L(R_0)$   in the limit $L\to \infty$ can be understood  from the fact that fractional strings of finite length carry an infinitely small fraction of the central charge of an infinitely long string.

At any finite value of $R_0$, the number of modes that can be excited on the long string is infinite as well, since the excitation energy $\epsilon_L$ goes to zero in the limit.
This balances $c_L(R_0)$ in such a way that the total number of degrees of freedom at any radius $R_0$ is still finite. This is manifested in the fractional string frame, where from  (\ref{cLR}) it follows that  $c_{R_0}(R_0)=A(R_0)/16\pi G_d \neq 0$ and  
\begin{equation}
\Delta_{R_0}-\frac{c_{R_0}(R_0)}{12}=0 \, .
\end{equation}
This implies that $\Delta_{R_0}$ corresponds to the scaling dimension of an infinite degree twist operator.

Minkowski space can also be arrived at by taking the $L\to \infty$ limit of  de Sitter space.
As will become clear in the next section, in this case one can understand  Minkowski space from the massless BTZ perspective.

 \subsection{De Sitter space}
 
In this section, we will argue that the microscopic quantum system relevant for sub-AdS$_d$ holography, as studied in detail in the section \ref{sec:sub-ads}, plays an equally important role in the microscopic description of the static patch of dS$_d$. 
In particular, our methods identify the static patch dS$_d$ as an excited state in that quantum system, in contrast to AdS below its curvature scale which was identified as the groundstate.

As explained in section \ref{sec:three-ex}, the de Sitter static patch metric  times  a transversal circle,
\begin{equation}\label{eq:metric-ds-circle}
ds^2=-\left(1- \frac{R^2}{L^2} \right)dt^2+\left(1- \frac{R^2}{L^2} \right)^{-1} dR^2+R^2d\Omega^2_{d-2}+\ell^2d\Phi^2,
\end{equation}
is Weyl equivalent to a Hawking-Page BTZ black hole times a transversal sphere
 \begin{align}\label{eq:metric-btz-long-string}
\begin{split}
 ds^2& =\frac{R^2}{L^2}\left[-\left ({r^{\,2}\over L^2}-1\right) dt^2  +\left ({r^{\,2}\over L^2}-1\right)^{-1} \! \!dr^{2} +r^{2} d\phi^{2} + L^2d\Omega^2_{d-2}\right], \\
\end{split}
\end{align}
where again $\Phi \equiv \Phi + 2\pi$ and  $\phi\equiv \phi +2\pi \ell/L$. 
The master formula \eqref{eq:metric-gen-conf-dim} reproduces the latter metric when $\Delta_{L}=c_L(R)/6$.

Let us start again for $R=L$.
The resulting BTZ metric arises holographically from an excited state in the CFT with conformal dimension $\Delta_L=c_L(L)/6$, as can for instance be verified by the Cardy formula \eqref{CHR-formula}:
\begin{equation}\label{eq:entropy-btz}
S= \frac{2\pi L}{4G_3}\frac{\ell}{L},
\end{equation}
This is the correct entropy for the Hawking-Page BTZ black hole with a conical defect of order $N$ \cite{deBoer:2010ac}.
The CFT state can be interpreted as a thermal gas of long strings at temperature $T \sim {1}/{L}$. Hence, the temperature is of the same order as the excitation energy $\epsilon_L$ of a long string. Our conjecture now suggests that the microscopic quantum system at the radial slice $R=L$ in dS$_d$, which coincides with the de Sitter horizon, should sit in an excited state as well. In fact, we like to interpret this state, similarly as in the CFT, as consisting of long strings, where each string typically carries only its lowest energy excitation mode. 
 
In  the gluing of $dS_d\times S^1$ and $BTZ\times S^{d-2}$, as explained in section  \ref{sec:three-ex}, we identify the horizon in the former spacetime with the horizon in the latter.
The entropy of the de Sitter space can then be understood as the entropy of the CFT$_2$ state.
Using the relations between Newton's constants in \eqref{eq:metric-btz-long-string} at the radial slice $R=L$,
$$
\frac{1}{G_3}=\frac{A(L)}{G_{d+1}},\qquad \frac{1}{G_d}=\frac{2\pi \ell}{G_{d+1}},
$$
one quickly finds that the BTZ entropy can be rewritten as: 
\begin{equation}\label{eq:desitter-entropy}
S =\frac{A(L)}{4 G_d}.
\end{equation}
Thus, we see that the Bekenstein-Hawking entropy for a $d$-dimensional de Sitter horizon can be reproduced from the Cardy formula in two-dimensional CFT.

An important point we should stress here is the reason for why the excitation of the sub-AdS$_d$ degrees of freedom in this case does not produce a Hawking-Page AdS$_d$ black hole.
This black hole would indeed have the same entropy as in \eqref{eq:desitter-entropy}, so what is it that enables us to distinguish them?
At the AdS scale, there is in fact nothing that distinguishes them, so to answer this question we must turn to the microscopic quantum system that describes sub-AdS$_d$ scales.
The crucial difference is that the CFT$_2$ state corresponding to the AdS$_d$ black hole has a constant conformal dimension, as we will come back to in detail in section \ref{sec:bhentropy}. 
On the other hand, the state corresponding to a sub-dS slice has $R_0$ dependent conformal dimension $\Delta_L=c_L(R_0)/6$, and therefore has a description in terms of the long strings in the $k^{d-1}$-fold symmetric product.
This is also clear from the master formula \eqref{eq:metric-gen-conf-dim}.
The dependence of the conformal dimension on $R_0$ expresses the fact that (part of) the excitations corresponding to de Sitter horizon are also present at sub-dS scales.
It is useful to move to a fractional string perspective, where we have:
\begin{equation}\label{eq:conf-dim-de-sitter-fract}
\Delta_{R_0} -  { c_{R_0}(R_0)\over 12} = \frac{R_0^2}{L^2} { c_{R_0} (R_0) \over 12} \,.
\end{equation}
This formula shows that only a part of the fractional strings at scale $R_0$ are excited, and in particular do not create a horizon, as should of course be the case for sub-dS.
It makes sense that only a fraction of the fractional strings are excited, since the de Sitter temperature $T\sim 1/L$ could only excite the longest strings with their lowest excitations.

As illustrated in Figure \ref{fig:penrose-ads-ds-inside-out}, the geometric perspective on sub-dS scales glues $dS_d\times S^1$ and $BTZ\times S^{d-2}$ by replacing the outer region $R_0\!<\!R\!<\!L$ in de Sitter with the region $r_h\!<\!\tilde{r}\!<\!R_0$ of the BTZ. 
The horizon size of this BTZ geometry is smaller than that of the Hawking-Page black hole.
This fact is expressed most clearly by the fractional string version of \eqref{eq:metric-btz-long-string}
\begin{equation}\label{eq:metric-btz-fractional-string}
  ds^2= \frac{R^2}{R_0^2} \left [ -\left ({\tilde r^{\,2}\over R_0^2}-\frac{R_0^2}{L^2}\right) dt^2  +\left ({\tilde r^{\,2}\over R_0^2}-\frac{R_0^2}{L^2}\right)^{-1} \! \!d\tilde r^{2} +\tilde r^{2} d\tilde\phi^{2} + R_0^2d\Omega^2_{d-2} \right] , 
\end{equation}
where for $R=R_0<L$ the horizon radius of the BTZ is given by $r_h=R^2_0/L<L$.
Note however that, as expected, the temperature of the black hole is not changed:
$$
T\sim \frac{r_h}{R_0^2}= \frac{1}{L}.
$$
Since in this case it is not the horizon of the smaller BTZ but the $\tilde{r}=R_0>r_h$ slice that is identified with the de Sitter slice $R=R_0$, one could wonder if the entropy of the BTZ can still be associated to the slice in de Sitter space.
However, in terms of the 2d CFT it is known that the entropy of a BTZ black hole is also contained in the states that live in the Hilbert space at higher energies than the black hole temperature \cite{Witten:1998zw}. Our conjecture then indeed suggests that the entropy of the smaller BTZ should also be associated to the sub-dS slice. 

The entropy of the smaller BTZ with the remaining angular deficit $\tilde{\phi}\equiv \tilde{\phi}+2\pi\ell/R_0$ in \eqref{eq:metric-btz-fractional-string} is given by:
\begin{equation}\label{eq:entropy-fract-btz}
S= \frac{2\pi R^2_0}{4G_3 L}\frac{\ell}{R_0}=\frac{A(R_0)}{4G_d}\frac{R_0}{L}.
\end{equation}
where we used the relation $1/G_3=A(R_0)/G_{d+1}$.
Hence, we see that  the entropy formula for sub-dS$_d$ scales with the volume instead of the area.
From the long string perspective, it is natural why only a fraction $(R_0/L)^{d-1}$ of the total de Sitter entropy arises at sub-dS scales. 
As we have discussed, de Sitter space corresponds to an excited state consisting of long strings at temperature $T\sim 1/L$. 
At sub-dS scales, $c_L(R_0)$ can be interpreted as $k^{d-2}$ fractions of long strings.
 It makes sense then that at the scale $R=R_0$ only a fraction of the energy and entropy associated to the long strings is accessible. 
If we apply the Cardy formula to the state $\Delta_{L}(R_0) =c_L(R_0)/6$, valid at least as long as $k^{d-1}\gg N$, we recover \eqref{eq:entropy-fract-btz}. We rewrite the entropy to make its volume dependence explicit as:
\begin{equation}\label{desitter-entropy-r0}
S=\frac{A(R_0)}{4G_d}\frac{R_0}{L} = \frac{V(R_0)}{V_0} \qquad \text{where} \qquad V_0 = \frac{4 G_d L }{d-1} \, . 
\end{equation}
Note that this entropy describes a volume law and only at the Hubble scale becomes the usual Bekenstein-Hawking entropy.
It hence seems natural to associate an entropy density to de Sitter space, which was advocated in \cite{Verlinde:2016toy}. However, the precise microscopic interpretation of this volume law remains an open question.

\subsection{Super-AdS scales revisited}
\label{sec:superadsrevisited}

In the previous sections we have given an interpretation of the microscopic holographic quantum system for sub-(A)dS$_d$ regions and flat space. To gain insight into sub-AdS$_d$ holography we used a conformal map to relate its foliation in holographic screens to the holographic screens of a family of super-AdS$_3$ screens. At this point, we have gained enough understanding of sub-AdS$_d$ to try to use the conformal map the other way around: we start with a sub-AdS$_3$ region and map it to a super-AdS$_d$ region. Even though the AdS/CFT correspondence already gives  a microscopic description of super-AdS$_d$ regions, we will argue that  this sub-AdS$_3$ perspective could still provide useful insights into the microscopic description of super-AdS$_d$ regions.
 
Let us start with the rescaled $AdS_3 \times S^{d-2}$ metric 
\begin{align}\label{eq:metric-sub-to-super}
\begin{split}
ds^2& =\frac{R_0^2}{L^2}\left[-\left ({r^{\,2}\over L^2}+1\right) dt^2  +\left ({r^{\,2}\over L^2}+1\right)^{-1} \! \!dr^{2} +r^{2} d\phi^{2} + L^2d\Omega^2_{d-2}\right] \, ,
\end{split}
\end{align}
where the angular coordinate has no deficit: $\phi\equiv \phi+2\pi$.
As should be familiar by now, this metric at slice $r_0=L^2/R_0$ is equivalent to AdS$_d$ at $R=R_0$ with a transversal circle of radius $L$.\footnote{Since we are now working at super-AdS$_d$ scales, we do not have to worry about a small size for the transversal $S^1$. However, it is perhaps not justified to ignore the transversal sphere at sub-AdS$_3$~scales.} We again glue the conformally equivalent spacetimes at $R=R_0$ and $r=r_0$, but now we take $R_0> L$ so that the sub-AdS$_3$ region is mapped to  a super-AdS$_d$ region.

The central charge for sub-AdS$_d$, as in \eqref{cLR}, together with the factor $R_0^2/L^2$ in front of the metric \eqref{eq:metric-sub-to-super} imply that the ``central charge'' associated to super-AdS$_d$ is given by
\begin{equation}\label{eq:super-adsd-central-charge-from-fract-string}
 \frac{c_L(R_0)}{12}=\frac{2\pi r_0}{16\pi G_3} \frac{r_0}{L}\left(\frac{R_0}{L}\right)^{d-1}  =\frac{A(R_0)}{16\pi G_d} \frac{L}{R_0}  \qquad \text{for} \qquad R_0>L  \, .
\end{equation}
This formula can be interpreted as the central charge of an $(R_0/L)^{d-1}$-fold symmetric product of $r_0$ sized fractional strings.
Note that this is a different quantity than the central charge of the CFT$_{d-1}$, given by formula  (\ref{centralcharge}).
Only in three dimensions the two expressions coincide and they  reproduce the   Brown-Henneaux formula for the central charge. The factor $L/R_0$ has an analogous interpretation as $R_0/L$ in the central charge at sub-AdS scales.
It expresses the fact that from the CFT$_2$ perspective, the degrees of freedom relevant at super-AdS$_d$ scales are fractions of long strings. 

Although $c_L(R_0)$ scales with $R_0^{d-3}$, the total number of quantum mechanical degrees of freedom is larger by a factor $R_0/L$. This is because the number of available modes on the long strings at energy scale $1/r_0=R_0/L^2$ is precisely $R_0/L$. As usual, the total amount of holographic degrees of freedom is reflected most clearly in the fractional string perspective:
\begin{align}
\begin{split} \label{NCsuperAdS}
\mathcal N    &=  \Delta_{R_0}-\frac{c_{R_0} (R_0)}{12}= \left (  \Delta_L-\frac{c_L (R_0)}{12}   \right) \frac{L}{R_0} \, ,  \\ 
\mathcal C   &=\frac{c_{R_0}(R_0)}{12} = \frac{c_L(R_0)}{12}  \frac{R_0}{L}   \, .  
\end{split}
\end{align}
Note that we found the same relations in (\ref{excnumber1}) for the  AdS$_3$/CFT$_2$ correspondence, but these equations hold for general $d$ and have a rather different interpretation, supplied by the sub-AdS$_3$ perspective. The  total number of degrees of freedom $\mathcal C$  is again given by an area law.

In conclusion, we see that a sub-AdS$_3$ perspective on super-AdS$_d$ identifies the degrees of freedom of the latter  with symmetric products of fractional strings. As we move outwards in AdS$_d$ the fractional string degrees of freedom become shorter and hence their excitation energy increases. 
In this way, the sub-AdS$_3$ perspective reproduces the usual UV-IR correspondence of super-AdS$_d$ holography.
We will use these results in section~\ref{sec:bhentropy}, when we discuss super-AdS$_d$ black holes.

\section{Physical implications}
\label{sec:physicalimplications}

As explained above, the long string sector of a    symmetric product CFT$_2$ gives  a detailed description of the   holographic degrees of freedom relevant for   non-AdS spacetimes. We now turn to a number of physical implications of this microscopic description. First, we will derive the Bekenstein-Hawking entropy for small and large AdS$_d$ black holes from a Cardy-like formula.
Moreover, from the CFT$_2$ point of view we will explain  why small black holes have a negative specific heat capacity and how  the Hawking-Page transition between small and large black holes  can be understood.
Finally, we will show that our long string perspective   reproduces the value of the vacuum energy for (A)dS spacetimes.

 \subsection{Black hole entropy and negative specific heat}
 \label{sec:bhentropy}
 
In this section we will apply our microscopic description   to small and large black holes in AdS$_d$.
We start with the case of small black holes, whose horizon size  $R_h$ is smaller than the AdS scale $L$.
For this situation, we consider the metric in the master formula \eqref{eq:metric-gen-conf-dim} for
\begin{equation} \label{eq:deltablackholemass}
\Delta_L=\frac{ML}{d-2} \, .
\end{equation}
Then, one may check that the AdS-Schwarzschild metric (with a transversal circle) follows after making performing the usual coordinate transformation $R=L^2/r$ and $\Phi=N\phi$:
\begin{equation}\label{eq:metric-ads-schw}
ds^2\!=\!-\! \left (\!1\!+\!{R^{\,2}\over L^2}\!-\! \frac{16\pi G_d M}{(d\!-\!2)\Omega_{d-2}R^{d-3}}\right)\!  dt^2  \!+\! \left (1\!+\!{R^{\,2}\over L^2}\! -\! \frac{16\pi G_d M}{(d\! -\! 2)\Omega_{d-2}R^{d-3}}\right)^{-1} \! \!\! \!\!dR^{2} \!+\! R^{2}d\Omega^2_{d- 2} \!+\! \ell^2 d\Phi^{2}  ,
\end{equation}
where $\Phi \equiv \Phi + 2\pi$. In particular, this implies that  the holographic screen at $R=R_0$ in AdS-Schwarzschild is equivalent to the $r_0=L^2/R_0$ screen in the metric \eqref{eq:metric-gen-conf-dim} for $\Delta_L$ as above.

In contrast to the previous cases, this time we do not have a clear interpretation of the AdS$_3$ metric for any value of $R$.
This is because for the particular value of $\Delta_L$ the AdS$_3$ metric between brackets in \eqref{eq:metric-gen-conf-dim} contains an $r^{(d-1)}$-dependent term.
This fact prohibits an analysis of AdS-Schwarzschild analogous to the previous cases.
However, we propose to overcome this difficulty by considering this metric only at a single constant slice $r=r_h$, for which the $g_{tt}$ component is zero.
This happens when  
\begin{equation}\label{eq:horizon-eqn-non-btz}
\Delta_L-\frac{c_{L}(R_h)}{12}=\frac{L^2}{R_h^2}\frac{c_{L}(R_h)}{12} \, ,
\end{equation}
where  we have used $R_h=L^2/r_h$.
We now interpret this slice as the horizon of an ordinary BTZ with horizon radius $r_h$, where we make use of the fact  that horizons are locally indistinguishable \cite{Jacobson:2003wv}.
In the AdS$_d$-Schwarzschild metric, this slice becomes of course precisely the horizon $R=R_h$ of the AdS$_d$ black hole.

Using a fractional string phenomenon we can  also express the relation above as:
\begin{equation}
\Delta_{R_h} - \frac{c_{R_h}(R_h)}{12} =\frac{c_{R_h}(R_h)}{12} \qquad \text{or} \qquad \mathcal N = \mathcal C \, .
\end{equation}
Therefore, from the AdS$_3$ point of view we can think of a black hole with horizon radius $R_h$ in AdS$_d$ as a thermal bath of $(R_h/\ell)^{d-2}$ fractional strings at temperature $T\sim 1/R_h$.
At the AdS$_3$ radial slice $r=r_h$ these are all available degrees of freedom, so it is very natural that a black hole arises in the AdS$_d$ frame.

The entropy of the AdS$_d$ black hole may now be computed from a CFT$_2$ perspective.
Indeed, applying the Cardy formula to the state in \eqref{eq:horizon-eqn-non-btz} yields
\begin{equation}\label{ads-schw-entropy-r0}
S=4\pi \sqrt{\frac{c_{L}(R_h)}{6}\left(\Delta_{L}-\frac{c_{L}(R_h)}{12}\right)}=\frac{A(R_h) }{4G_d} \, .
\end{equation}
The CFT$_2$ perspective tells us that, at the level of counting holographic degrees of freedom and their excitations, the Bekenstein-Hawking formula is a Cardy formula.
This perspective may explain the appearance of a Virasoro algebra and corresponding Cardy formula found in \cite{Majhi:2011ws,Majhi:2012tf}.  

Next we discuss large AdS$_d$ black holes, with horizon size $R_h >L$.
For this situation we use the results from section \ref{sec:superadsrevisited}, where we related    super-AdS$_d$ holography to sub-AdS$_3$ physics. By inserting the relations  (\ref{NCsuperAdS}) for $\mathcal N$ and $\mathcal C$ into (\ref{metricdef}) we find the following metric for sub-AdS$_3$ scales
\begin{equation}\label{eq:metric-gen-conf-dim-3}
ds^2 =\frac{R^2}{L^2}\left[-\left(\!1\!-\!\frac{r^2}{L^2}\frac{\Delta_{L}-\frac{c_L(R)}{12}}{\frac{c_L(R)}{12}}\right)\!dt^2+\left(\!1\!-\!\frac{r^2}{L^2}\frac{\Delta_{L}-\frac{c_L(R)}{12}}{\frac{c_L(R)}{12}}\right)^{-1}\!dr^2\!+\!r^2d\phi^2\!+\!L^2d\Omega^2_{d-2}\right].
\end{equation}
This turns into the AdS-Schwarzschild metric if one inserts formula (\ref{eq:deltablackholemass}) for $\Delta_L$ and equation \eqref{eq:super-adsd-central-charge-from-fract-string} for $c_L(R)$.
The horizon  equation now becomes
\begin{equation}
\Delta_L-\frac{ c_{L}(R_h)}{12}=\frac{R^2_h}{L^2}\frac{c_{L}(R_h)}{12} \, .
\end{equation}
Applying the Cardy formula to this state can easily be seen to reproduce the Bekenstein-Hawking entropy for a super-AdS black hole.
This provides an explanation for the fact that the Bekenstein-Hawking entropy for AdS$_d$ black holes can be written as a Cardy-like formula for CFT$_{d-1}$ \cite{Verlinde:2000wg}.
Indeed, as for the small black holes discussed above, it shows that the AdS black hole entropy at this level of discussion \emph{is} a Cardy formula.
This result also indicates the potential usefulness of a sub-AdS$_3$ perspective on super-AdS$_d$ scales, even though the latter should be completely accessible by the CFT$_{d-1}$.

Finally, we comment on the Hawking-Page transition between small and large AdS black holes \cite{Hawking:1982dh}. 
The fact that a super-AdS$_d$ black hole has positive specific heat can be understood from the AdS$_3$ perspective in the following way. 
As we increase the size $R_h$ of the AdS$_d$ black hole, we are decreasing the radius $r_h$ in sub-AdS$_3$.
At the same time, we are adding degrees of freedom since the metric is multiplied by an ever growing factor $(R_h/L)^2>1$.
Decreasing $r_h$ at sub-AdS scales is in the direction of the UV.
In other words, the excitations have to become of larger energy since they should fit on smaller strings.
Therefore, we see that the black hole heats up as we increase its size, and hence it has a positive specific heat.
Of course, this was already well understood without referring to AdS$_3$.
The positive specific heat namely originates from the fact that the CFT$_{d-1}$ energy scale is proportional to the AdS$_d$ radial coordinate here and the number of degrees of freedom is proportional to the area of the radial slice.

On the other hand, for sub-AdS$_d$ black holes, as we make the horizon size $R_h$ smaller, we are increasing   $r_h$ in the super-AdS$_3$ perspective and therefore the black hole heats up.
The decrease in the number of degrees of freedom in AdS$_3$, even though we are moving outwards to larger $r_h$, is due to the factor $(R_h/L)^2<1$ that multiplies the metric.
Thus, the black hole heats up as we make it smaller, which establishes the negative specific heat.
There is no clear CFT$_{d-1}$ interpretation of this fact, so this is one of the main new insights from our AdS$_3$ perspective on sub-AdS$_d$~scales.

To conclude, the crucial aspect that leads to the negative specific heat is the fact that the UV-IR correspondence is reversed at sub-AdS scales.
Turning this around, one could have viewed the negative specific heat of small black holes as an important clue for the reversal of the UV-IR correspondence.

\subsection{Vacuum energy of (A)dS}
\label{sec:vacenergy}

Finally, we give an interpretation of the vacuum energy of (A)dS$_d$ spacetime from our CFT$_2$ perspective.
The vacuum energy density of (A)dS is related to the cosmological constant through $\rho_{\text{vac}} = \Lambda/8\pi G_d$ and is hence set by the IR scale. 
However, from quantum field theory one expects that the vacuum energy is instead sensitive to the UV cut-off, for example the Planck scale.
In our interpretation, the UV and IR scale correspond, respectively, to the short and long string length. 
 We will argue below that the long string phenomenon precisely explains why the vacuum energy is set by the long string scale instead of the short string scale.

The vacuum energy of $d$-dimensional (A)dS   contained in a spacelike region with volume $V(R)$ is given by 
\begin{equation}  \label{vacuumenergy}
E^{\text{(A)dS}}_{\text{vac}}= \pm \frac{(d-1)(d-2) }{16 \pi G_d L^2} V(R) \qquad \text{with} \qquad V(R) = \frac{\Omega_{d-2}R^{d-1}}{d-1} \, .
\end{equation}
The vacuum energy of AdS is negative, whereas that of dS space is positive. This expression is   obtained by multiplying the energy density $\rho_{\text{vac}} = \Lambda/8\pi G_d$ associated with the cosmological constant $\Lambda = - (d-1)(d-2)/2L^2$ with what appears to be the flat  volume of the spherical region.\footnote{A more accurate explanation of the simple expression for $V(R)$ is that in addition to describing the proper volume it incorporates the redshift factor.}  We now want to a give a microscopic interpretation of this formula  in terms of the long strings discussed in the previous section. 

In the CFT$_2$ the energy of a quantum state can be computed by multiplying the excitation number with the excitation energy of a degree of freedom. 
For long strings, this reads:
\begin{equation} \label{vacuumenergyL}
E=\left(\Delta_L-\frac{c_L(R)}{12}\right) \epsilon_L \, . 
\end{equation}
The subtraction $- c_L(R)/12$ is due to the negative Casimir energy of the CFT on a circle. 
The scaling dimension $\Delta_L$, the central charge $c_L(R)$ and the excitation energy $\epsilon_L=(d-2)/L$ are all labeled by the long string scale $L$.   
The vacuum energy of (A)dS$_d$ then follows from the expression above by imposing   specific values for the scaling dimension, $\Delta_L = 0$~(AdS) and $\Delta_L = c_L(R)/6$ (dS), i.e. 
\begin{equation}
E_{\text{vac}}^{\text{(A)dS}} = \pm \frac{c_L (R)}{12} \, \epsilon_L \, .
\end{equation}
 By inserting the values of $c_{L}(R)$ and $\epsilon_L$ into the formula above one recovers the correct vacuum energy (\ref{vacuumenergy}). This formula tells us that we can interpret the negative vacuum energy of AdS as a Casimir energy, which is what it corresponds to in the CFT$_2$. The vacuum energy of de Sitter space can be attributed to the excitations of   the lowest energy states available in the system: the long strings of size $L$.

Alternatively, if one assumes the vacuum energy is determined by the UV  or short string  degrees of freedom, one would have instead computed:
\begin{equation}
E_{\text{vac}}^{\text{UV}}   =  \pm \frac{c_\ell (R)}{12} \,   \epsilon_\ell   \, .
 \end{equation}
 By inserting the values for the short string central charge \eqref{eq:centralcharge-shortstring} and  the  excitation energy $\epsilon_\ell = (d-2)/\ell$, we find for their vacuum energy:
\begin{equation}  \label{eq:uvvacenergy}
E_{\text{vac}}^{\text{UV}}   = \pm \frac{(d-1)(d-2) }{16 \pi G_d \ell^2} V(R)   \, .
\end{equation}
This is off by a factor $1/N^2 = \ell^2/L^2$ from the true vacuum energy of (A)dS space.
The long string phenomenon precisely explains why the vacuum energy associated to long strings is $N^2$ times lower than the vacuum energy associated to short strings. It namely decreases both the central charge and the energy gap by a factor of $N$.
Thus, the identification of the long strings as the correct holographic degrees of freedom   provides a natural explanation of the value of the vacuum energy.

\section{Conclusion and discussion}
\label{sec:conclusion}

In this paper we have   proposed a new approach to  holography for non-AdS spacetimes, in particular: AdS below its curvature radius, Minkowski, de Sitter and AdS-Schwarzschild. Before summarizing our main findings, we comment on the more general lessons for non-AdS holography. First of all, we would like to make a distinction between 
  holography as manifested by the AdS/CFT correspondence and holography in general, which is only constrained by the holographic principle. The original principle states that the number of degrees of freedom in quantum gravity is bounded by the Bekenstein-Hawking formula. On the other hand, AdS/CFT is a much stronger statement,
in which the holographic degrees of freedom of quantum gravity are identified as part of a local quantum field theory in one dimension less.
We do expect (and indeed assume) the holographic principle to hold for more general spacetimes, motivated by the standard black hole  arguments \cite{tHooft:1993dmi,Susskind:1994vu,Bousso:1999xy}. However, there are indications that it is unlikely to expect a local quantum field theory dual to gravity in non-AdS spacetimes. 

One can already arrive at this conclusion for sub-AdS scales via the following reasoning. Let us assume a lattice regularization of the boundary CFT where each lattice site contains a number of degrees of freedom proportional to the central charge \cite{Susskind:1998dq}. The fact that the central charge is related to the area at the AdS radius  (cf. equation   (\ref{centralcharge}))  implies that one is left with a single lattice site   as one holographically renormalizes up to the AdS scale.
This means that the effective theory on a holographic screen at the AdS radius is completely delocalized, and can be described by  a matrix quantum mechanics.\footnote{The same conclusion is reached when considering the flat space   limit of AdS/CFT \cite{Susskind:1998vk}. In addition, generalizations of the Ryu-Takayanagi proposal to more general spacetimes suggest a non-local dual description of flat space and de Sitter space \cite{Li:2010dr,Shiba:2013jja,Miyaji:2015yva,Sanches:2016sxy,Nomura:2016ikr,Nomura:2017fyh,Nomura:2018kji}. }
It is not obvious how to further renormalize the quantum theory to probe the interior of a single AdS region. One expects, however,  that the degrees of freedom ``inside the matrix'' should play a role in the holographic description of sub-AdS regions  \cite{Susskind:1998dq,Berenstein:2005aa}. 
The reasoning also shows that one has to be careful in applying the Ryu-Takayanagi formula to sub-AdS scales, since   it typically assumes a spatial factorization of the holographic Hilbert space. For recent discussions of this issue, see for instance \cite{Balasubramanian:2014sra,Yang:2015uoa,Nomura:2018kji}. 
It is expected that similar conclusions apply to flat space holography, which should be described by the $L \to \infty$ limit of sub-AdS holography, and de Sitter static patch holography, which is connected to flat space holography via the same limit.

Even though a local quantum field theory dual may not exist for non-AdS spacetimes, we do assume that there exists a dual quantum mechanical theory which can be associated to a holographic screen. 
In this paper we have proposed to study some general features of such quantum mechanical  theories for non-AdS geometries. These features are captured by three quantities associated to a holographic screen at radius $R$: the number of degrees of freedom $\mathcal C$, the excitation number $\mathcal N$ and the excitation energy $ \epsilon$. 
We have given a more refined interpretation of these quantities in terms of a twisted sector of a symmetric product CFT$_2$. This is achieved through a conformal map between the non-AdS geometries and locally AdS$_3$ spacetimes.  In the CFT$_2$ language, the holographic degrees of freedom are interpreted as (fractions of) long strings. 

The qualitative picture that arises is as follows. The symmetric product theory introduces a number of short degrees of freedom that scales with the volume of the non-AdS spacetime. However, the long string phenomenon, by gluing together short degrees of freedom, reduces the number of degrees of freedom to an area law, consistent with the Bekenstein-Hawking formula. We have also seen that the degrees of freedom on holographic screens at distance scales smaller than the (A)dS radius should be thought of as fractions of long strings. 
This perspective suggests that the long string degrees of freedom extend into the bulk, instead of being localized on a holographic screen. This stands in contrast with the holographic degrees of freedom that describe large AdS regions which, as suggested by holographic renormalization, are localized on their associated screens.
   
For all of the non-AdS spacetimes we studied, one of the main conclusions that arises from our proposal is that the number of degrees of freedom in the microscopic holographic theory increases towards the IR in the bulk.  
This follows from the reversal of the UV-IR relation and the holographic principle.
The familiar UV-IR correspondence thus appears to be a special feature of the holographic description of AdS space at scales larger than its curvature radius.  Our results suggest that in general the UV and IR in the spacetime geometry and in the microscopic theory are in sync.
A similar point of view has appeared in \cite{Anninos:2011af}, where the term `worldline holography' was coined (see also the review \cite{Anninos:2012qw} and their recent work \cite{Anninos:2017hhn}). This term refers to the fact that the UV observer is placed at the center of spacetime, instead of at a boundary as in AdS holography. These two observers are related by the conformal map that inverts the radius and exchanges the UV with IR.

The UV-IR correspondence in AdS/CFT is in line with the Wilsonian intuition for a quantum field theory and thus explains why the microscopic holographic theory can be described by a QFT.  On the other hand, we have argued that the general features of non-AdS holography are naturally accommodated for by symmetric products and the long string phenomenon. 
Our results then suggest that the holographic dual of non-AdS spacetimes cannot be described by (Wilsonian) quantum field theories, but  should rather be thought of as quantum mechanical systems that exhibit the long string phenomenon. 

An example of such a model is given  by matrix quantum mechanics. For instance, in the BFSS matrix model \cite{Banks:1996vh} a long string phenomenon was observed to play a role \cite{Dijkgraaf:1997vv}, inspired by the results of \cite{Taylor:1996ik}. In particular, the $N\to \infty$ limit of the matrix model corresponds to a large symmetric product, and in the far IR the long strings are the only surviving degrees of freedom. We propose that one possible way in which a smaller UV Hilbert space can be embedded in a larger IR Hilbert space, is to identify the UV degrees of freedom as excitations on fractional strings and view these excited fractional strings as bound states of the lowest energy excitations on long strings that live in the  far IR.

As we explained, the symmetric products and long string phenomenon also provide a natural framework to understand the negative specific heat of small AdS black holes.
Moreover, using a sub-AdS$_3$ perspective on super-AdS$_d$ scales, we have shown how the Hawking-Page transition between positive and negative specific heat black holes in AdS$_d$ can be understood in the CFT$_2$ language.
Even though the positive specific heat also follows from the CFT$_{d-1}$ description \cite{Witten:1998zw}, the thermal state in a strongly coupled CFT$_{\!d\!-\!1}$ is by no means a simple object to study.
It would be interesting to see whether our sub-AdS$_3$ perspective and the associated CFT$_2$ language could provide a new avenue to study strongly coupled higher dimensional CFTs.
One result we obtained in this direction is a way to understand the appearance of a Cardy-like formula in $(d\!-\!1)$-dimensional (holographic) CFTs, that describes the entropy of a $d$-dimensional AdS black hole \cite{Verlinde:2000wg,Majhi:2011ws,Majhi:2012tf}.
Namely, our conformal map relates the AdS$_d$ black hole entropy to the   entropy of a BTZ black hole, which can be derived from the Cardy formula.

We should note that the negative specific heat of small $AdS_5\times S^5$ black holes has already been studied from the $\mathcal N=4$ SYM theory in \cite{Asplund:2008xd}. 
In this paper, it is argued that a sub-matrix of the large $N$ matrix could provide a description of such ten-dimensional black holes, including the negative specific heat.
It would be   interesting to understand whether our set-up could be generalized to cover or be embedded in an $AdS_5\times S^5$ geometry.
As argued in their paper, the sub-matrix forms an essentially isolated system that consists of a dense gas of strings. 
It does not thermalize with its environment, i.e. does not spread on the $S^5$. 
Concerning the embedding, our small $S^1$, on which we also have a dense gas of strings for small AdS black holes, could be the effective geometry seen by such a localized, confined system on $S^5$.

More generally, an open question is whether the  geometries in this paper can be embedded in string  or M-theory. 
A particularly interesting case that we leave for future study is the MSW string \cite{Maldacena:1997de}.
This string has a near-horizon geometry of the form $AdS_3\times S^2$. Applying our conformal map to a BTZ geometry in this set-up would lead to a  $dS_4\times S^1$ spacetime. 
The microscopic quantum description of the latter spacetime could be related to the $D0$-$D4$ quiver quantum mechanics theories studied by \cite{Gaiotto:2004ij}.  It is plausible that an extension of Matrix theory to $d=4$ requires the inclusion of transversal fivebranes \cite{Berkooz:1996is} and precisely leads to such a quiver QM description. 

Finally, we have argued that de Sitter space must be regarded as an excited state of the microscopic  holographic quantum system.\footnote{This is also suggested by the generalization of the Ryu-Takayanagi proposal to cosmological spacetimes, where the  holographic entanglement entropy obeys a volume law on the   holographic screen \cite{Nomura:2016ikr}.} In this sense it is similar to the BTZ spacetime to which it is conformally equivalent. 
This explains in particular the fact that the de Sitter entropy can be recovered from a Cardy formula in two-dimensional CFT.
From its description as a thermal bath of long strings at the Gibbons-Hawking temperature $T\sim 1/L$, we moreover showed that   the  vacuum energy   of de Sitter space can be reproduced from the energy carried by the long strings.

 \section*{Acknowledgments}

We woud like to thank Dionysios Anninos, Jan de Boer,  and Herman Verlinde for interesting   discussions, and Juan Maldacena for his suggestion to study   the conformal map from the embedding formalism. This work is supported by the Spinoza grant and the DITP consortium, which are both funded by the Dutch science organization NWO.

\appendix

\section{Weyl equivalent spacetimes from an embedding   formalism}

\label{embedding}

In this paper we have studied three cases of conformally equivalent spacetimes:
\begin{align}\label{eq:conformal-equivalences-quotiented-p=3}
 AdS_{d}\times S^{1}  &\cong \text{conical} \, \,  AdS_{3}    \times S^{d-2}\, ,    \nonumber  \\
Mink_d\times S^1  & \cong \text{massless} \,\,     {BTZ}    \times S^{d-2}\, ,     \\
 dS_{d}\times S^1 & \cong \text{Hawking-Page} \, \,     {BTZ}     \times S^{d-2} \, . \nonumber
\end{align} 
The $(d+1)$-dimensional spacetimes on the left and right hand side are  Weyl equivalent with the specific conformal factor  $\Omega=L/r = R/L.$ Notice that the spacetimes on the right hand side are all    discrete quotients of (patches of) pure  AdS$_3$. They can  be obtained by orbifolding AdS$_3$ by an elliptic, parabolic and hyperbolic element of the isometry group, respectively \cite{Banados:1992gq}. On the left hand side, the quotient acts on the one-dimensional space. Since taking the quotients commutes with the Weyl transformation we could also consider the conformal equivalence of the unquotiented spaces. These are in fact easier to understand and can be generalized to any dimension $p$ as follows: 
\begin{align}
 AdS_{d}\times S^{p-2} & \cong AdS_{p}\times S^{d-2},    \nonumber   \\
Mink_d\times \mathbb{R}^{p-2}  & \cong \text{Poincar\'{e}-}AdS_{p}\times S^{d-2}, \\
 dS_{d}\times \mathbb{H}^{p-2}   & \cong \text{Rindler-}AdS_{p}\times S^{d-2}.  \nonumber
\end{align} 
Note that when $\mathbb{H}^1\cong \mathbb{R}$ is quotiented by a boost, one obtains the $S^1$ in \eqref{eq:conformal-equivalences-quotiented-p=3}.
In this appendix we will   explain in detail how the conformal equivalence of these spacetimes can be understood from the embedding space perspective.

\paragraph{Embedding space formalism.} The $(p+d-2)$-dimensional spacetimes above are   special in the sense that they are conformally flat. Now any $D$-dimensional conformally flat spacetime  can be embedded in $\mathbb R^{2,D-2}$, where the metric signature of the original spacetime is $(-,  +,   \dots ,+)$. This can be seen as follows. To begin with,   Minkowski spacetime can be obtained  as a section of the light cone through the origin of $\mathbb R^{2,D-2}$. The light cone equation is
\begin{equation} 
X \cdot X = -X^2_{-1}-X^2_0+X^2_1+\ldots +X^2_{D}=0.
\end{equation}
Here $X_A$ are the standard flat coordinates on $\mathbb R^{2,D-2}$. The embedding space naturally induces a metric on the light cone section. The Poincar\'e section $X_{-1} + X_{D}=1$, for example, leads to the standard   metric on Minkowski spacetime. Under the coordinate transformation, $  X_A = \Omega (x) \tilde X_A$, the induced metric becomes
\begin{equation}   
d   s^{2} =d {X}\cdot d {X}=(\Omega d\tilde X+\tilde Xd\Omega)^2=\Omega^2d\tilde X\cdot d\tilde X=\Omega^2 d\tilde s^2 \, .
\end{equation}
Here the light cone properties $\tilde X \cdot d \tilde X =0$ and $\tilde X \cdot \tilde X =0$ were used in the third equality. This means that the induced metrics on two different light cone sections are related by a Weyl transformation. 
Thus, any spacetime which is conformally flat can be embedded in $\mathbb R^{2,D-2}.$

 \paragraph{Global AdS.}
The first class of conformally equivalent spacetimes is given by
\begin{equation}\label{eq:global-ads-weyl-equivalence}
  AdS_{d}\times S^{p-2} \cong AdS_{p}\times S^{d-2} \, .
\end{equation}
Both spacetimes can be obtained as a section  of the light cone in the embedding space $\mathbb{R}^{2,p+d-2}$
\begin{equation}\label{eq:null-cone}
-X^2_{-1}-X^2_0+X^2_1+\ldots +X^2_{p+d-2}=0 \, .
\end{equation}
The scaling symmetry $X_A \to \lambda X_A$ of this equation can be fixed in multiple ways. Each choice corresponds to a different section of the light cone and realizes a different conformally flat spacetime. For instance,
\begin{equation}\label{eq:global-ads-scale-inv-fix1}
\underbrace{-\tilde X^2_{-1}-\tilde X^2_0}_{=\:- r^2- L^2}+\underbrace{\tilde X^2_1+\ldots+ \tilde X^2_{p-1}}_{=\:r^2}+\underbrace{\tilde X^2_{p}+\ldots+\tilde X^2_{p+d-2}}_{=\:L^2}=0  
\end{equation}
corresponds to  $AdS_{p}\times S^{d-2}$, where $r$ parametrizes the radial direction in AdS$_p$  and $L$ is the size of the sphere $S^{d-2}$.
On the other hand,
\begin{equation}\label{eq:global-ads-scale-inv-fix2}
\underbrace{- X^2_{-1}- X^2_0}_{=\:-L^2 -R^2}+\underbrace{ X^2_1+\ldots+  X^2_{p-1}}_{=\:L^2}+\underbrace{  X^2_{p}+\ldots+ X^2_{p+d-2}}_{=\:R^2}=0
\end{equation}
leads to $AdS_{d}\times S^{p-2}$, where $R$ is the radial coordinate in $\text{AdS}_d$. It is straightforward to see that the induced metrics on the two sections   are related by the Weyl transformation
\begin{equation}   \label{weylrelation}
d  s^2_{(d, p-2)} = \Omega^2  d \tilde s^2_{(p,d-2)}   \qquad \text{with} \qquad \Omega = \frac{L}{r} = \frac{R}{L} \, .
\end{equation}
The explicit expressions for the conformally equivalent metrics are  
\begin{equation}\label{eq:ads-sphere-pq-metric}
 d \tilde s^2_{(p,d-2)}=-\left(\frac{r^2}{L^2}+1\right)dt^2+\left( \frac{r^2}{L^2} +1\right)^{-1} dr^2+r^2d\Omega^2_{p-2}+L^2d\Omega^2_{d-2} \, , 
\end{equation}
and
\begin{equation}\label{eq:ads-sphere-qp-metric}
d  s^2_{(d, p-2)}=-\left(1 + \frac{R^2}{L^2}\right)dt^2+ \left(1+\frac{R^2}{L^2}\right)^{-1} dR^2+R^2d\Omega^2_{d-2}+L^2d\Omega^2_{p-2} \, . 
\end{equation}

\paragraph{Poincar\'e patch.}
The second class of Weyl equivalent geometries is
\begin{equation}\label{eq:ads-poinc-weyl-equivalence}
 \mathbb{R}^{1,d+p-2}  \cong \text{Poincar\'e-}AdS_{p}\times S^{d-2} \, .
\end{equation}
The Poincar\'e patch   can be understood as a flat foliation of AdS$_p$ with leaves  $\mathbb{R}^{1,p-2}$.
The choice of coordinates on the light  cone  (\ref{eq:null-cone}) that leads to the geometry on  the right hand side is
\begin{equation}\label{eq:ads-poinc-scale-inv-fix1}
\underbrace{-\tilde X^2_{-1}+\tilde X^2_{p-1}}_{=\: - \frac{r^2}{L^2}x^{\mu}x_{\mu}- L^2 }\underbrace{-\tilde X^2_{0}+\tilde X^2_1+\ldots+ \tilde X^2_{p-2}}_{=\:\frac{r^2}{L^2}x^{\mu}x_{\mu}}+\underbrace{\tilde X^2_{p}+\ldots+\tilde X^2_{p+d-2}}_{=\:L^2}=0 \, ,
\end{equation}
where $r$ parametrizes the radial direction in the Poincar\'e patch and $x^\mu$ represent flat coordinates on $\mathbb{R}^{1,p-2}$.   The section can also simply be described by  $\tilde X_{-1} + \tilde X_{p-1} = L$.

Alternatively, we can fix the scale invariance to obtain $\mathbb{R}^{1,d+p-2}$ through
\begin{equation}\label{eq:ads-poinc-scale-inv-fix2}
\underbrace{-  X^2_{-1}+  X^2_{p-1}}_{=\: -x^{\mu}x_{\mu}-R^2}\underbrace{-  X^2_{0}+  X^2_1+\ldots+   X^2_{p-2}}_{=\:x^{\mu}x_{\mu}}+\underbrace{  X^2_{p}+\ldots+  X^2_{p+d-2}}_{=\:R^2}=0 \, .
\end{equation}
Both of these sections are so-called Poincar\'{e} sections and the latter one is described by  the equation $  X_{-1} +   X_{p-1} = R$. One can  easily verify that the induced metrics on these two sections are related by (\ref{weylrelation}), and they  explicitly  take the form
\begin{equation}\label{eq:poinc-path-metric}
d\tilde s^2_{(p, d-2)}=- \frac{r^2}{L^2} dt^2  +  \frac{L^2}{r^2}dr^2+\frac{r^2}{L^2} d \vec x^2_{p-2}  +  L^2d\Omega^2_{d-2} \, ,
\end{equation}
and
\begin{equation}\label{eq:mink-metric}
d {s}^2_{(d,p-2)}=- dt^2+dR^2 +R^2d\Omega^2_{d-2}+ d \vec x^2_{p-2}\, .
\end{equation}

\paragraph{AdS-Rindler patch.}
The third class  is given by 
\begin{equation}\label{eq:ads-rindler-weyl-equivalence}
  dS_{d}\times \mathbb{H}^{p-2} \cong  \text{Rindler-}AdS_{p}\times S^{d-2} \, ,
\end{equation}
where  $\mathbb{H}^{p-2}$ denotes $(p-2)$-dimensional hyperbolic space.
In this case, we use the fact that time slices of the AdS-Rindler patch are foliated by   hyperbolic space $\mathbb{H}^{p-2}$.
The scale fixing of the null cone that leads to  the geometry on the right hand side is
\begin{equation}\label{eq:ads-rindler-scale-inv-fix1}
\underbrace{-\tilde X^2_0+\tilde X^2_1}_{=\:-L^2+r^2}\underbrace{-\tilde X^2_{-1}+\tilde X^2_2+\ldots+\tilde X^2_{p-1}}_{=\:-r^2}+\underbrace{\tilde X^2_{p}+\ldots+\tilde X^2_{p+d-2}}_{=\:L^2}=0 \, ,
\end{equation}
where $r$ parametrizes the radial direction in the Rindler wedge.
Furthermore,  one can  obtain $dS_{d}\times \mathbb{H}^{p-2}$ by fixing the   coordinates on the light cone as follows
\begin{equation}\label{eq:ads-rindler-scale-inv-fix2}
\underbrace{-  X^2_0+  X^2_1}_{=\:-R^2+L^2}\underbrace{-  X^2_{-1}+  X^2_2+\ldots+   X^2_{p-1}}_{=\:-L^2}+\underbrace{  X^2_{p}+\ldots+  X^2_{p+d-2}}_{=\:R^2}=0 \, .
\end{equation}
Again, one easily verifies that this implies that the induced metrics on the sections are Weyl equivalent with conformal factor $\Omega =L /r=R/L$.
Explicitly, the induced metrics are given by
\begin{equation}\label{eq:ads-rin-sphere-metric}
d\tilde s^2_{(p, d-2)}=-\left(\frac{r^2}{L^2}-1\right)dt^2+ \left( \frac{r^2}{L^2}-1\right)^{-1} dr^2+r^2\left(du^2+\sinh^2(u)d\Omega^2_{p-3}\right)+L^2d\Omega^2_{d-2} \, ,
\end{equation}
and
\begin{equation}\label{eq:ds-hyperboloid-metric}
d {s}^2_{(d,p-2)}=-\left(1-\frac{R^2}{L^2}\right)dt^2+ \left(1-\frac{R^2}{L^2}\right)^{-1} dR^2+R^2d\Omega^2_{d-2}+L^2\left(du^2+\sinh^2(u)d\Omega^2_{p-3}\right) \, .
\end{equation}
As mentioned above and used in the main text, in the case that $p=3$ we may compactify $\mathbb{H}^1$ by taking a discrete quotient by a boost. On the right hand side, the AdS-Rindler side, this same identification produces a BTZ black hole.

\bibliographystyle{JHEP}
\bibliography{manus}

\providecommand{\href}[2]{#2}\begingroup\raggedright\begin{thebibliography}{10}

\bibitem{Maldacena:1997re}
J.~M. Maldacena, \emph{{The Large N limit of superconformal field theories and
  supergravity}}, {\emph{Int. J. Theor. Phys.} {\bf 38} (1999) 1113--1133},
  [\href{https://arxiv.org/abs/hep-th/9711200}{{\tt hep-th/9711200}}].

\bibitem{Witten:1998qj}
E.~Witten, \emph{{Anti-de Sitter space and holography}},
  \href{http://dx.doi.org/10.4310/ATMP.1998.v2.n2.a2}{\emph{Adv. Theor. Math.
  Phys.} {\bf 2} (1998) 253--291},
  [\href{https://arxiv.org/abs/hep-th/9802150}{{\tt hep-th/9802150}}].

\bibitem{Aharony:1999ti}
O.~Aharony, S.~S. Gubser, J.~M. Maldacena, H.~Ooguri and Y.~Oz, \emph{{Large N
  field theories, string theory and gravity}},
  \href{http://dx.doi.org/10.1016/S0370-1573(99)00083-6}{\emph{Phys. Rept.}
  {\bf 323} (2000) 183--386}, [\href{https://arxiv.org/abs/hep-th/9905111}{{\tt
  hep-th/9905111}}].

\bibitem{Henningson:1998gx}
M.~Henningson and K.~Skenderis, \emph{{The Holographic Weyl anomaly}},
  \href{http://dx.doi.org/10.1088/1126-6708/1998/07/023}{\emph{JHEP} {\bf 07}
  (1998) 023}, [\href{https://arxiv.org/abs/hep-th/9806087}{{\tt
  hep-th/9806087}}].

\bibitem{Akhmedov:1998vf}
E.~T. Akhmedov, \emph{{A Remark on the AdS / CFT correspondence and the
  renormalization group flow}},
  \href{http://dx.doi.org/10.1016/S0370-2693(98)01270-2}{\emph{Phys. Lett.}
  {\bf B442} (1998) 152--158},
  [\href{https://arxiv.org/abs/hep-th/9806217}{{\tt hep-th/9806217}}].

\bibitem{Alvarez:1998wr}
E.~Alvarez and C.~Gomez, \emph{{Geometric holography, the renormalization group
  and the c theorem}},
  \href{http://dx.doi.org/10.1016/S0550-3213(98)00752-4}{\emph{Nucl. Phys.}
  {\bf B541} (1999) 441--460},
  [\href{https://arxiv.org/abs/hep-th/9807226}{{\tt hep-th/9807226}}].

\bibitem{Balasubramanian:1999jd}
V.~Balasubramanian and P.~Kraus, \emph{{Space-time and the holographic
  renormalization group}},
  \href{http://dx.doi.org/10.1103/PhysRevLett.83.3605}{\emph{Phys. Rev. Lett.}
  {\bf 83} (1999) 3605--3608},
  [\href{https://arxiv.org/abs/hep-th/9903190}{{\tt hep-th/9903190}}].

\bibitem{Skenderis:1999mm}
K.~Skenderis and P.~K. Townsend, \emph{{Gravitational stability and
  renormalization group flow}},
  \href{http://dx.doi.org/10.1016/S0370-2693(99)01212-5}{\emph{Phys. Lett.}
  {\bf B468} (1999) 46--51}, [\href{https://arxiv.org/abs/hep-th/9909070}{{\tt
  hep-th/9909070}}].

\bibitem{deBoer:1999tgo}
J.~de~Boer, E.~P. Verlinde and H.~L. Verlinde, \emph{{On the holographic
  renormalization group}},
  \href{http://dx.doi.org/10.1088/1126-6708/2000/08/003}{\emph{JHEP} {\bf 08}
  (2000) 003}, [\href{https://arxiv.org/abs/hep-th/9912012}{{\tt
  hep-th/9912012}}].

\bibitem{Bekenstein:1973ur}
J.~D. Bekenstein, \emph{{Black holes and entropy}},
  \href{http://dx.doi.org/10.1103/PhysRevD.7.2333}{\emph{Phys. Rev.} {\bf D7}
  (1973) 2333--2346}.

\bibitem{Hawking:1974sw}
S.~W. Hawking, \emph{{Particle Creation by Black Holes}},
  \href{http://dx.doi.org/10.1007/BF02345020}{\emph{Commun. Math. Phys.} {\bf
  43} (1975) 199--220}.

\bibitem{tHooft:1993dmi}
G.~'t~Hooft, \emph{{Dimensional reduction in quantum gravity}}, {\emph{Conf.
  Proc.} {\bf C930308} (1993) 284--296},
  [\href{https://arxiv.org/abs/gr-qc/9310026}{{\tt gr-qc/9310026}}].

\bibitem{Susskind:1994vu}
L.~Susskind, \emph{{The World as a hologram}},
  \href{http://dx.doi.org/10.1063/1.531249}{\emph{J. Math. Phys.} {\bf 36}
  (1995) 6377--6396}, [\href{https://arxiv.org/abs/hep-th/9409089}{{\tt
  hep-th/9409089}}].

\bibitem{Bousso:1999xy}
R.~Bousso, \emph{{A Covariant entropy conjecture}},
  \href{http://dx.doi.org/10.1088/1126-6708/1999/07/004}{\emph{JHEP} {\bf 07}
  (1999) 004}, [\href{https://arxiv.org/abs/hep-th/9905177}{{\tt
  hep-th/9905177}}].

\bibitem{Bousso:2002ju}
R.~Bousso, \emph{{The Holographic principle}},
  \href{http://dx.doi.org/10.1103/RevModPhys.74.825}{\emph{Rev. Mod. Phys.}
  {\bf 74} (2002) 825--874}, [\href{https://arxiv.org/abs/hep-th/0203101}{{\tt
  hep-th/0203101}}].

\bibitem{Susskind:1998dq}
L.~Susskind and E.~Witten, \emph{{The Holographic bound in anti-de Sitter
  space}},  \href{https://arxiv.org/abs/hep-th/9805114}{{\tt hep-th/9805114}}.

\bibitem{Strominger:1996sh}
A.~Strominger and C.~Vafa, \emph{{Microscopic origin of the Bekenstein-Hawking
  entropy}}, \href{http://dx.doi.org/10.1016/0370-2693(96)00345-0}{\emph{Phys.
  Lett.} {\bf B379} (1996) 99--104},
  [\href{https://arxiv.org/abs/hep-th/9601029}{{\tt hep-th/9601029}}].

\bibitem{Maldacena:1996ds}
J.~M. Maldacena and L.~Susskind, \emph{{D-branes and fat black holes}},
  \href{http://dx.doi.org/10.1016/0550-3213(96)00323-9}{\emph{Nucl. Phys.} {\bf
  B475} (1996) 679--690}, [\href{https://arxiv.org/abs/hep-th/9604042}{{\tt
  hep-th/9604042}}].

\bibitem{Dijkgraaf:1996xw}
R.~Dijkgraaf, G.~W. Moore, E.~P. Verlinde and H.~L. Verlinde, \emph{{Elliptic
  genera of symmetric products and second quantized strings}},
  \href{http://dx.doi.org/10.1007/s002200050087}{\emph{Commun. Math. Phys.}
  {\bf 185} (1997) 197--209}, [\href{https://arxiv.org/abs/hep-th/9608096}{{\tt
  hep-th/9608096}}].

\bibitem{Myers:2010tj}
R.~C. Myers and A.~Sinha, \emph{{Holographic c-theorems in arbitrary
  dimensions}}, \href{http://dx.doi.org/10.1007/JHEP01(2011)125}{\emph{JHEP}
  {\bf 01} (2011) 125}, [\href{https://arxiv.org/abs/hep-th/1011.5819}{{\tt
  hep-th/1011.5819}}].

\bibitem{Osborn:1993cr}
H.~Osborn and A.~C. Petkou, \emph{{Implications of conformal invariance in
  field theories for general dimensions}},
  \href{http://dx.doi.org/10.1006/aphy.1994.1045}{\emph{Annals Phys.} {\bf 231}
  (1994) 311--362}, [\href{https://arxiv.org/abs/hep-th/9307010}{{\tt
  hep-th/9307010}}].

\bibitem{Brown:1986nw}
J.~D. Brown and M.~Henneaux, \emph{{Central Charges in the Canonical
  Realization of Asymptotic Symmetries: An Example from Three-Dimensional
  Gravity}}, \href{http://dx.doi.org/10.1007/BF01211590}{\emph{Commun. Math.
  Phys.} {\bf 104} (1986) 207--226}.

\bibitem{Banados:1992wn}
M.~Ba\~nados, C.~Teitelboim and J.~Zanelli, \emph{{The Black hole in
  three-dimensional space-time}},
  \href{http://dx.doi.org/10.1103/PhysRevLett.69.1849}{\emph{Phys. Rev. Lett.}
  {\bf 69} (1992) 1849--1851},
  [\href{https://arxiv.org/abs/hep-th/9204099}{{\tt hep-th/9204099}}].

\bibitem{Kraus:2006wn}
P.~Kraus, \emph{{Lectures on black holes and the $AdS_3 / CFT_2$
  correspondence}}, {\emph{Lect. Notes Phys.} {\bf 755} (2008) 193--247},
  [\href{https://arxiv.org/abs/hep-th/0609074}{{\tt hep-th/0609074}}].

\bibitem{Strominger:1997eq}
A.~Strominger, \emph{{Black hole entropy from near horizon microstates}},
  \href{http://dx.doi.org/10.1088/1126-6708/1998/02/009}{\emph{JHEP} {\bf 02}
  (1998) 009}, [\href{https://arxiv.org/abs/hep-th/9712251}{{\tt
  hep-th/9712251}}].

\bibitem{Banks:2011av}
T.~Banks, \emph{{Holographic Space-Time: The Takeaway}},
  \href{https://arxiv.org/abs/hep-th/1109.2435}{{\tt hep-th/1109.2435}}.

\bibitem{Banks:2013qpa}
T.~Banks, \emph{{Lectures on Holographic Space Time}},
  \href{https://arxiv.org/abs/hep-th/1311.0755}{{\tt hep-th/1311.0755}}.

\bibitem{Brewin}
L.~Brewin, \emph{{A Simple expression for the ADM mass}},
  \href{http://dx.doi.org/10.1007/s10714-007-0403-9}{\emph{Gen. Rel. Grav.}
  {\bf 39} (2007) 521--528}, [\href{https://arxiv.org/abs/gr-qc/0609079}{{\tt
  gr-qc/0609079}}].

\bibitem{Bousso:1999cb}
R.~Bousso, \emph{{Holography in general space-times}},
  \href{http://dx.doi.org/10.1088/1126-6708/1999/06/028}{\emph{JHEP} {\bf 06}
  (1999) 028}, [\href{https://arxiv.org/abs/hep-th/9906022}{{\tt
  hep-th/9906022}}].

\bibitem{Ryu:2006bv}
S.~Ryu and T.~Takayanagi, \emph{{Holographic derivation of entanglement entropy
  from AdS/CFT}},
  \href{http://dx.doi.org/10.1103/PhysRevLett.96.181602}{\emph{Phys. Rev.
  Lett.} {\bf 96} (2006) 181602},
  [\href{https://arxiv.org/abs/hep-th/0603001}{{\tt hep-th/0603001}}].

\bibitem{Ryu:2006ef}
S.~Ryu and T.~Takayanagi, \emph{{Aspects of Holographic Entanglement Entropy}},
  \href{http://dx.doi.org/10.1088/1126-6708/2006/08/045}{\emph{JHEP} {\bf 08}
  (2006) 045}, [\href{https://arxiv.org/abs/hep-th/0605073}{{\tt
  hep-th/0605073}}].

\bibitem{Li:2010dr}
W.~Li and T.~Takayanagi, \emph{{Holography and Entanglement in Flat
  Spacetime}},
  \href{http://dx.doi.org/10.1103/PhysRevLett.106.141301}{\emph{Phys. Rev.
  Lett.} {\bf 106} (2011) 141301}, [\href{https://arxiv.org/abs/1010.3700}{{\tt
  1010.3700}}].

\bibitem{Shiba:2013jja}
N.~Shiba and T.~Takayanagi, \emph{{Volume Law for the Entanglement Entropy in
  Non-local QFTs}},
  \href{http://dx.doi.org/10.1007/JHEP02(2014)033}{\emph{JHEP} {\bf 02} (2014)
  033}, [\href{https://arxiv.org/abs/1311.1643}{{\tt 1311.1643}}].

\bibitem{Miyaji:2015yva}
M.~Miyaji and T.~Takayanagi, \emph{{Surface/State Correspondence as a
  Generalized Holography}},
  \href{http://dx.doi.org/10.1093/ptep/ptv089}{\emph{PTEP} {\bf 2015} (2015)
  073B03}, [\href{https://arxiv.org/abs/1503.03542}{{\tt 1503.03542}}].

\bibitem{Sanches:2016sxy}
F.~Sanches and S.~J. Weinberg, \emph{{Holographic entanglement entropy
  conjecture for general spacetimes}},
  \href{http://dx.doi.org/10.1103/PhysRevD.94.084034}{\emph{Phys. Rev.} {\bf
  D94} (2016) 084034}, [\href{https://arxiv.org/abs/1603.05250}{{\tt
  1603.05250}}].

\bibitem{Nomura:2016ikr}
Y.~Nomura, N.~Salzetta, F.~Sanches and S.~J. Weinberg, \emph{{Toward a
  Holographic Theory for General Spacetimes}},
  \href{http://dx.doi.org/10.1103/PhysRevD.95.086002}{\emph{Phys. Rev.} {\bf
  D95} (2017) 086002}, [\href{https://arxiv.org/abs/1611.02702}{{\tt
  1611.02702}}].

\bibitem{Nomura:2017fyh}
Y.~Nomura, P.~Rath and N.~Salzetta, \emph{{Spacetime from Unentanglement}},
  \href{https://arxiv.org/abs/1711.05263}{{\tt 1711.05263}}.

\bibitem{Nomura:2018kji}
Y.~Nomura, P.~Rath and N.~Salzetta, \emph{{Pulling the Boundary into the
  Bulk}},  \href{https://arxiv.org/abs/1805.00523}{{\tt 1805.00523}}.

\bibitem{Anninos:2011af}
D.~Anninos, S.~A. Hartnoll and D.~M. Hofman, \emph{{Static Patch Solipsism:
  Conformal Symmetry of the de Sitter Worldline}},
  \href{http://dx.doi.org/10.1088/0264-9381/29/7/075002}{\emph{Class. Quant.
  Grav.} {\bf 29} (2012) 075002},
  [\href{https://arxiv.org/abs/hep-th/1109.4942}{{\tt hep-th/1109.4942}}].

\bibitem{Hubeny:2009rc}
V.~E. Hubeny, D.~Marolf and M.~Rangamani, \emph{{Hawking radiation from AdS
  black holes}},
  \href{http://dx.doi.org/10.1088/0264-9381/27/9/095018}{\emph{Class. Quant.
  Grav.} {\bf 27} (2010) 095018},
  [\href{https://arxiv.org/abs/hep-th/0911.4144}{{\tt hep-th/0911.4144}}].

\bibitem{Martinec:1998wm}
E.~J. Martinec, \emph{{Conformal field theory, geometry, and entropy}},
  \href{https://arxiv.org/abs/hep-th/9809021}{{\tt hep-th/9809021}}.

\bibitem{Balasubramanian:2014sra}
V.~Balasubramanian, B.~D. Chowdhury, B.~Czech and J.~de~Boer,
  \emph{{Entwinement and the emergence of spacetime}},
  \href{http://dx.doi.org/10.1007/JHEP01(2015)048}{\emph{JHEP} {\bf 01} (2015)
  048}, [\href{https://arxiv.org/abs/1406.5859}{{\tt 1406.5859}}].

\bibitem{deBoer:2010ac}
J.~de~Boer, M.~M. Sheikh-Jabbari and J.~Simon, \emph{{Near Horizon Limits of
  Massless BTZ and Their CFT Duals}},
  \href{http://dx.doi.org/10.1088/0264-9381/28/17/175012}{\emph{Class. Quant.
  Grav.} {\bf 28} (2011) 175012},
  [\href{https://arxiv.org/abs/hep-th/1011.1897}{{\tt hep-th/1011.1897}}].

\bibitem{Witten:1998zw}
E.~Witten, \emph{{Anti-de Sitter space, thermal phase transition, and
  confinement in gauge theories}},
  \href{http://dx.doi.org/10.4310/ATMP.1998.v2.n3.a3}{\emph{Adv. Theor. Math.
  Phys.} {\bf 2} (1998) 505--532},
  [\href{https://arxiv.org/abs/hep-th/9803131}{{\tt hep-th/9803131}}].

\bibitem{Verlinde:2016toy}
E.~P. Verlinde, \emph{{Emergent Gravity and the Dark Universe}},
  \href{http://dx.doi.org/10.21468/SciPostPhys.2.3.016}{\emph{SciPost Phys.}
  {\bf 2} (2017) 016}, [\href{https://arxiv.org/abs/1611.02269}{{\tt
  1611.02269}}].

\bibitem{Jacobson:2003wv}
T.~Jacobson and R.~Parentani, \emph{{Horizon entropy}},
  \href{http://dx.doi.org/10.1023/A:1023785123428}{\emph{Found. Phys.} {\bf 33}
  (2003) 323--348}, [\href{https://arxiv.org/abs/gr-qc/0302099}{{\tt
  gr-qc/0302099}}].

\bibitem{Majhi:2011ws}
B.~R. Majhi and T.~Padmanabhan, \emph{{Noether Current, Horizon Virasoro
  Algebra and Entropy}},
  \href{http://dx.doi.org/10.1103/PhysRevD.85.084040}{\emph{Phys. Rev.} {\bf
  D85} (2012) 084040}, [\href{https://arxiv.org/abs/gr-qc/1111.1809}{{\tt
  gr-qc/1111.1809}}].

\bibitem{Majhi:2012tf}
B.~R. Majhi and T.~Padmanabhan, \emph{{Noether current from the surface term of
  gravitational action, Virasoro algebra and horizon entropy}},
  \href{http://dx.doi.org/10.1103/PhysRevD.86.101501}{\emph{Phys. Rev.} {\bf
  D86} (2012) 101501}, [\href{https://arxiv.org/abs/gr-qc/1204.1422}{{\tt
  gr-qc/1204.1422}}].

\bibitem{Verlinde:2000wg}
E.~P. Verlinde, \emph{{On the holographic principle in a radiation dominated
  universe}},  \href{https://arxiv.org/abs/hep-th/0008140}{{\tt
  hep-th/0008140}}.

\bibitem{Hawking:1982dh}
S.~W. Hawking and D.~N. Page, \emph{{Thermodynamics of Black Holes in anti-De
  Sitter Space}}, \href{http://dx.doi.org/10.1007/BF01208266}{\emph{Commun.
  Math. Phys.} {\bf 87} (1983) 577}.

\bibitem{Susskind:1998vk}
L.~Susskind, \emph{{Holography in the flat space limit}},
  \href{http://dx.doi.org/10.1063/1.1301570}{\emph{AIP Conf. Proc.} {\bf 493}
  (1999) 98--112}, [\href{https://arxiv.org/abs/hep-th/9901079}{{\tt
  hep-th/9901079}}].

\bibitem{Berenstein:2005aa}
D.~Berenstein, \emph{{Large N BPS states and emergent quantum gravity}},
  \href{http://dx.doi.org/10.1088/1126-6708/2006/01/125}{\emph{JHEP} {\bf 01}
  (2006) 125}, [\href{https://arxiv.org/abs/hep-th/0507203}{{\tt
  hep-th/0507203}}].

\bibitem{Yang:2015uoa}
Z.~Yang, P.~Hayden and X.-L. Qi, \emph{{Bidirectional holographic codes and
  sub-AdS locality}},
  \href{http://dx.doi.org/10.1007/JHEP01(2016)175}{\emph{JHEP} {\bf 01} (2016)
  175}, [\href{https://arxiv.org/abs/1510.03784}{{\tt 1510.03784}}].

\bibitem{Anninos:2012qw}
D.~Anninos, \emph{{De Sitter Musings}},
  \href{http://dx.doi.org/10.1142/S0217751X1230013X}{\emph{Int. J. Mod. Phys.}
  {\bf A27} (2012) 1230013}, [\href{https://arxiv.org/abs/1205.3855}{{\tt
  1205.3855}}].

\bibitem{Anninos:2017hhn}
D.~Anninos and D.~M. Hofman, \emph{{Infrared Realization of dS$_2$ in
  AdS$_2$}},  \href{https://arxiv.org/abs/hep-th/1703.04622}{{\tt
  hep-th/1703.04622}}.

\bibitem{Banks:1996vh}
T.~Banks, W.~Fischler, S.~H. Shenker and L.~Susskind, \emph{{M theory as a
  matrix model: A Conjecture}},
  \href{http://dx.doi.org/10.1103/PhysRevD.55.5112}{\emph{Phys. Rev.} {\bf D55}
  (1997) 5112--5128}, [\href{https://arxiv.org/abs/hep-th/9610043}{{\tt
  hep-th/9610043}}].

\bibitem{Dijkgraaf:1997vv}
R.~Dijkgraaf, E.~P. Verlinde and H.~L. Verlinde, \emph{{Matrix string theory}},
  \href{http://dx.doi.org/10.1016/S0550-3213(97)00326-X}{\emph{Nucl. Phys.}
  {\bf B500} (1997) 43--61}, [\href{https://arxiv.org/abs/hep-th/9703030}{{\tt
  hep-th/9703030}}].

\bibitem{Taylor:1996ik}
W.~Taylor, \emph{{D-brane field theory on compact spaces}},
  \href{http://dx.doi.org/10.1016/S0370-2693(97)00033-6}{\emph{Phys. Lett.}
  {\bf B394} (1997) 283--287},
  [\href{https://arxiv.org/abs/hep-th/9611042}{{\tt hep-th/9611042}}].

\bibitem{Asplund:2008xd}
C.~T. Asplund and D.~Berenstein, \emph{{Small AdS black holes from SYM}},
  \href{http://dx.doi.org/10.1016/j.physletb.2009.02.043}{\emph{Phys. Lett.}
  {\bf B673} (2009) 264--267}, [\href{https://arxiv.org/abs/0809.0712}{{\tt
  0809.0712}}].

\bibitem{Maldacena:1997de}
J.~M. Maldacena, A.~Strominger and E.~Witten, \emph{{Black hole entropy in M
  theory}}, \href{http://dx.doi.org/10.1088/1126-6708/1997/12/002}{\emph{JHEP}
  {\bf 12} (1997) 002}, [\href{https://arxiv.org/abs/hep-th/9711053}{{\tt
  hep-th/9711053}}].

\bibitem{Gaiotto:2004ij}
D.~Gaiotto, A.~Strominger and X.~Yin, \emph{{Superconformal black hole quantum
  mechanics}},
  \href{http://dx.doi.org/10.1088/1126-6708/2005/11/017}{\emph{JHEP} {\bf 11}
  (2005) 017}, [\href{https://arxiv.org/abs/hep-th/0412322}{{\tt
  hep-th/0412322}}].

\bibitem{Berkooz:1996is}
M.~Berkooz and M.~R. Douglas, \emph{{Five-branes in M(atrix) theory}},
  \href{http://dx.doi.org/10.1016/S0370-2693(97)00014-2}{\emph{Phys. Lett.}
  {\bf B395} (1997) 196--202},
  [\href{https://arxiv.org/abs/hep-th/9610236}{{\tt hep-th/9610236}}].

\bibitem{Banados:1992gq}
M.~Ba\~{n}ados, M.~Henneaux, C.~Teitelboim and J.~Zanelli, \emph{{Geometry of
  the (2+1) black hole}}, \href{http://dx.doi.org/10.1103/PhysRevD.48.1506,
  10.1103/PhysRevD.88.069902}{\emph{Phys. Rev.} {\bf D48} (1993) 1506--1525},
  [\href{https://arxiv.org/abs/gr-qc/9302012}{{\tt gr-qc/9302012}}].

\end{thebibliography}\endgroup

\end{document}